\documentclass[12pt]{article}
\usepackage{savesym}
\usepackage{cite}
\usepackage{wasysym}
\usepackage{amscd}
\usepackage{graphicx}
\usepackage{pifont}
\usepackage{float}
\usepackage{subfig}
\usepackage[all]{xy}
\usepackage{multicol}
\usepackage{amsfonts}
\usepackage{picture}
\usepackage{amssymb}
\savesymbol{iint}
\savesymbol{iiint}
\usepackage{amsmath}
\usepackage{amsthm}

\restoresymbol{TXF}{iint}
\restoresymbol{TXF}{iiint}
\usepackage{color}
\usepackage{clay}
\usepackage{hyperref}
\usepackage{slashed}
\usepackage{tikz,pgfplots}
\usepgfplotslibrary{fillbetween}
\hypersetup{colorlinks=true}
\hypersetup{linkcolor=black}
\hypersetup{citecolor=black}
\hypersetup{urlcolor=black}
\usepackage{setspace}
\usepackage{multirow}
\usepackage{wrapfig}
\usepackage{verbatim}
\numberwithin{equation}{section}

\newcommand{\beq}{\begin{equation}}
\newcommand{\eeq}{\end{equation}}
\newcommand{\nn}{\nonumber\\} 
\newcommand{\bea}{\begin{eqnarray}}
\newcommand{\ea}{\end{eqnarray}}
\newcommand{\barr}{\begin{array}}
\newcommand{\earr}{\end{array}}
\newcommand{\lb}{{\langle}}
\newcommand{\rb}{{\rangle}}
\newcommand{\beajm}{\begin{eqnarray}}
  \newcommand{\eeajm}{\end{eqnarray}}

\newcommand{\OO}{\mathcal{O}}
\newcommand{\D}{\Delta}
\newcommand{\spc}{\hspace{1mm}}

\def\la{\label}
\def\nref#1{(\ref{#1})}
\newcommand{\be}{\begin{equation}}
  
  \newcommand{\ee}{\end{equation}}
 \newcommand{\half}{{1\over 2}}

\begin{document}

\date{October, 2017}

\institution{IAS}{\centerline{${}^{1}$School of Natural Sciences, Institute for Advanced Study, Princeton, NJ, USA}}
\institution{Princeton}{\centerline{${}^{2}$Physics Department, Princeton University, Princeton, NJ, USA}}

\title{ Bounds on OPE Coefficients from \\Interference Effects in the Conformal Collider }

\authors{Clay C\'{o}rdova\worksat{\IAS}\footnote{e-mail: {\tt claycordova@ias.edu}}, Juan Maldacena\worksat{\IAS}\footnote{e-mail: {\tt malda@ias.edu}}, and Gustavo J. Turiaci\worksat{\Princeton}\footnote{e-mail: {\tt turiaci@princeton.edu}}}

\abstract{  

We apply the average null energy condition to obtain upper bounds on the three-point function coefficients of stress tensors and a scalar operator,  $\langle TT {\cal O } \rangle,$  in
general CFTs. We also constrain the gravitational anomaly of $U(1)$ currents in four-dimensional CFTs, which are encoded in three-point functions of the form $\langle TT J \rangle$. 
In theories with a large $N$ AdS dual we translate these bounds into constraints on the coefficient of a 
 higher derivative bulk  term of the form $\int \phi\hspace{.5mm} W^2 $. 
We speculate that these bounds also apply in de-Sitter.  In this case our results constrain inflationary observables, such as the amplitude for chiral gravity waves 
that originate from higher derivative terms in the Lagrangian of the form  
 $\phi \hspace{.5mm}W W^*$. 
 }

\maketitle

\setcounter{tocdepth}{2}
\begingroup
\setstretch{1}
\tableofcontents
\endgroup

\newpage
\section{Introduction}

In this paper we investigate some implications of the average null energy condition in conformal field theories. 
We consider the conformal collider physics experiment discussed in \cite{Hofman:2008ar}. In that setup, we produce a localized 
excitation by acting with a smeared operator near the origin of spacetime. Then we measure the energy flux at infinity per unit angle. 
Requiring that the energy flux is positive imposes constraints on the three-point function coefficients. 
 This method was used to constrain three-point functions of the stress tensor in   \cite{Hofman:2008ar,deBoer:2009pn,Buchel:2009sk}. 

In this paper we   use this same method to constrain the three-point functions of two stress tensors and another  operator $\langle TT {\cal O } \rangle $. 
The new idea consists of creating the initial state by a linear combination of a stress tensor operator and the  operator ${\cal O}$. 
The three-point function $\langle TT {\cal O} \rangle $ appears as a kind of interference term in the expression for the energy.
 Requiring that the  total contribution   to the 
energy flux is positive imposes a non-trivial upper bound on the absolute magnitude of this three-point correlator.
We apply these ideas to general scalar operators ${\cal O}$ as well as conserved currents with spin one, $J$, where we use it to put bounds on the 
gravitational anomaly in $d=4$ CFTs. 
Because the bound arises from   quantum mechanical interference effects, these bounds are stronger than those obtained in states created by a single primary local operator and its descendants (though the resulting bounds involve more OPE coefficients).


This energy flux at infinity is given by an integral of the stress tensor. On the boundary of Minkowski space this integral is simply the average null energy
 ${\cal E} = \int dx^- T_{--} $. We review this in section \ref{CCmeth}. 
 Physically, we expect that this energy should be positive for all angles. 
Recently, the average null energy condition was proven using entanglement entropy methods \cite{Faulkner:2016mzt} as well as reflection positivity euclidean methods
\cite{Hartman:2016lgu}. 
 When we create a localized state using the stress tensor, this energy distribution is completely determined by the three-point function of the stress tensor. Two 
 of the insertions correspond to the insertions creating the state in the bra and the ket. The third corresponds to the one measuring the energy flux at infinity. 
The resulting bounds   could also be obtained by requiring standard reflection positivity 
of the euclidean theory  \cite{Hartman:2016dxc,Hofman:2016awc}.  However, the conformal collider calculations provide an efficient way to extract the results.

 %


One of our main results is a sum rule constraining the OPE coefficients of scalar primary operators $\mathcal{O}$ with the energy-momentum tensor $T$.  In spacetime dimensions $d\geq 4$ there is a single OPE coefficient controlling the $\langle TT\mathcal{O}\rangle$ three-point function.  We find that this data is constrained as 
\begin{equation}
\sum_{\text{Scalar Primaries}~\mathcal{O}_{i}} \hspace{-.1in}|C_{TT\mathcal{O}_{i}}|^{2}~ f(\Delta_{i})\leq N_{B}~, \label{boundintro}
\end{equation}
where $N_{B}$ is one of the three OPE coefficients in $\langle TTT\rangle$ (the one occurring in a theory of free bosons), and the non-negative function $f(\Delta)$ is given explicitly by 
\begin{equation} \la{Fvalue}
f(\Delta)= \frac{(d-1)^3 d \pi ^{2 d} \Gamma \left(\frac{d}{2}\right) \Gamma (d+1) \Gamma (\Delta ) \Gamma \left(\Delta -\frac{d-2}{2}\right)}{(d-2)^2 \Gamma \left(\frac{\Delta }{2}+2\right)^4 \Gamma \left(\frac{d+\Delta }{2}\right)^2 \Gamma \left(d-\frac{\Delta }{2}\right)^2}~.
\end{equation}
This function arises by doing the integrals involved in smearing the operator as well as in computing the energy flux. 
We derive this bound in detail in section \ref{TTOdg4}, and discuss some simple physical consequences such as its interpretation in free field theories, large $N$ holographic systems, and general implications for the asymptotics of OPE coefficients.

In section \ref{TTOd3} we consider analogous results in spacetime dimension three.  This case is special because the three-point functions of interest admit both parity preserving and parity violating structures.  The bounds we find generalize those recently obtained in \cite{Chowdhury:2017vel}.  We apply our results to large $N$ Chern-Simons matter theories, and further use them to obtain predictions on OPE coefficients $C_{TT {\cal O} }$ for scalars in the Ising model using the recent results of the conformal bootstrap \cite{Dymarsky:2017yzx}.  For instance, we find that operator $\varepsilon$ has an OPE coefficient constrained as
\begin{equation}
|C_{TT\varepsilon}|\leq 1.751 |C_{TT:\phi^{2}:}|~,
\end{equation} 
where the right-hand side is the value in the free scalar theory based on the field $\phi$.

In section \ref{TTJ} we consider bounds in four-dimensional CFTs with a global symmetry current $J$.  We apply the same techniques to obtain universal constraints on the gravitational anomaly of the current $J.$ 

In section \ref{AdSsection} we show that the $\langle TT {\cal O } \rangle$ correlator can be generated from 
a gravity theory in $AdS_{d+1}$ through a higher derivative term, $ \int \phi W^2$, in the bulk effective action. 
We match the coefficient of this term to the $C_{TT{\cal O}}$ coefficient in the boundary theory 
by performing the same collider experiment in the bulk, where it involves  propagation through a shock wave. One interesting feature of this presentation is that the resulting bound is independent of the mass of $\phi.$ Thus, the $\Delta$ dependence of \eqref{boundintro} is purely kinematic and results from translating the boundary three-point function coefficient to a bulk interaction.  We use our $AdS$ presentation to show that $\alpha'$ corrections satisfy the bound. 

In section \ref{dSsection} we extrapolate the bounds we obtained in $AdS$ to  ``quasi bounds'' on the 
coefficients of the effective action in de Sitter space. We call them ``quasi-bounds'' 
because,   unfortunately,  for de-Sitter we do not know how to prove a sharp bound.    We can think of these  as
 a good indication for where the bulk effective theory should break down. 
We apply these ``quasi-bounds'' to constrain the amplitude of chiral gravity waves, and to constrain the violations of
the inflationary ``consistency condition''  for the two-point function. Both of these effect arise from 
higher curvature couplings of the form $\phi W^2 $ or $\phi W W^*$. 

In the appendices we include more explicit derivations of the material in the main sections. 

\section{ ANEC and the Conformal Collider} 
\la{CCmeth}

\subsection{The Average Null Energy Condition}

The null energy condition is a central assumption in many classical theorems of general relativity.  These results allow us to exclude unphysical spacetimes where causality violation, naked singularities, or other physical pathologies occur \cite{Fuller:1962zza}.

If we move beyond classical field theory, these results appear to be in doubt.  Quantum effects lead to fluctuations that prohibit any local operator from having a positive expectation value in every state \cite{Epstein:1965zza}.  (We review these ideas in appendix \ref{secnoloc}.) In particular the local energy density and other components of the energy-momentum tensor have negative expectation value in some states. 

Deeper investigation reveals a potential resolution.  While components of the energy-momentum tensor are pointwise non-positive, a weaker hypothesis, the so-called average null energy condition, is often sufficient to enforce causal behavior \cite{Friedman:1993ty}.  This condition states that the integral along a complete null geodesic of the null energy density is a positive definite operator
\begin{equation}
\mathcal{E}=\int_{-\infty}^{\infty} dx^{-}~T_{--} \geq 0~. \label{ANECintro}
\end{equation}

Recently there has been significant interest in understanding the average null energy condition \eqref{ANECintro} in the context of local quantum field theories.  In \cite{Hartman:2016lgu}, an argument was given establishing \eqref{ANECintro} in conformal field theories by examining the constraints of causality on the light-cone operator product expansion.  In  \cite{Faulkner:2016mzt}, an alternative argument was given linking the average null energy operator to entanglement entropy, then establishing positivity using strong subadditivity.  These information theoretic methods have also been extended to obtain new inequalities strengthening \eqref{ANECintro} \cite{Balakrishnan:2017bjg}.

Given that  the average null energy in quantum field theory is now a theorem, 
 it is interesting to take it as input and use it to constrain conformal field theory data. 

\subsection{The Conformal Collider}
\label{ccreview}

An efficient way to extract consequences of the average null energy condition in CFTs is to use the conformal collider setup of\cite{Hofman:2008ar}. 
  This technique is closely related to deep inelastic scattering experiments in conformal field theory \cite{Hofman:2008ar, Komargodski:2016gci}.  As we review, in the context of AdS/CFT these bounds arise from demanding causality of the bulk theory in a shockwave background.

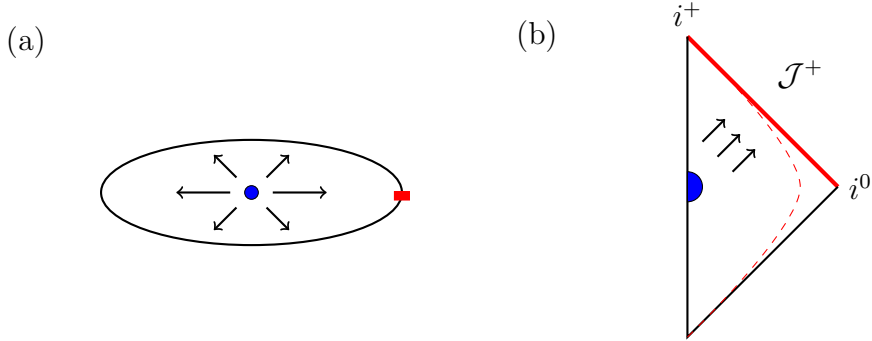
\begin{figure}[h]
\begin{center}
\begin{tikzpicture}[scale=1,rotate=0]
\draw[thick] (0,0) ellipse (2cm and 0.7cm);
\draw[thin, black, fill=blue] (0,0) circle (.5ex);
\draw[thick, ->] (-0.2,-0.2) -- (-0.5,-0.5) ;
\draw[thick, ->] (-0.2,0.2) -- (-0.5,0.5) ;
\draw[thick, ->] (0.2,0.2) -- (0.5,0.5) ;
\draw[thick, ->] (-0.2828,0) -- (-1,0) ;
\draw[thick, ->] (0.2828,0) -- (1,0) ;
\draw[thick, ->] (0.2,-0.2) -- (0.5,-0.5) ;
\draw[red,fill = red] (2cm-0.1cm,0cm-0.1cm) rectangle (2cm+0.1cm,0cm+0.01cm);
\node at (0,-1.8) {};
\node at (-3,2) {(a)};
\end{tikzpicture}
\hspace{1cm}
\begin{tikzpicture}[scale=1,rotate=0]
\draw[ultra thick,red] (0,2) -- (2,0) ;
\node at (0,2.3) {$i^+$};
\draw[thick] (2,0) -- (0,-2) -- (0,2);
\node at (2.3,0) {$i^0$};
\draw[thick, ->] (0.4,0.4) -- (0.7,0.7) ;
\draw[thick, ->] (0.2+0.4,-0.2+0.4) -- (0.2+0.7,-0.2+0.7) ;
\draw[thick, ->] (-0.2+0.4,0.2+0.4) -- (-0.2+0.7,0.2+0.7) ;
\node at (1.5,1.5) {$\mathcal{J}^+$};
\draw[thin, black , fill=blue] (0,0.2) arc(90:-90:0.2) -- cycle ;
\draw [red,dashed] plot [smooth] coordinates {(0,-2) (1.5,0) (0,2)};
\node at (-2,2) {(b)};
\end{tikzpicture}
\caption{ In the conformal collider experiment (a),  the energy created by a localized excitation (blue) is measured far away by a calorimeter (red). (b) For a CFT, this is equivalent to measuring the energy at null infinity $\mathcal{J}^+$. }
\label{FigureCollider}
\end{center}
\end{figure}

The specific physical problem of interest is to create a disturbance in a conformal field theory and then to measure the correlation of energy deposited at various angles at future null infinity (see Figure \ref{FigureCollider}). 

The states in which we measure the energy are obtained by acting with local operators $\mathcal{O}(x)$ on the Lorentzian vacuum $|0\rangle$. We further give these states definite timelike momentum $q$.\footnote{For technical reasons it is sometimes useful to create a localized wavepacket instead of an exact momentum eigenstate.  This subtlety will not affect our discussion.}  Thus we examine the state
\begin{equation}
|\mathcal{O}(q,\lambda)\rangle = \mathcal{N}\int d^{d}x ~ e^{-iqt} ~ \lambda \cdot \mathcal{O}(x)|0\rangle~, \label{statedef}
\end{equation}
where $\lambda$ is a polarization tensor accounting for the possible spin of $\mathcal{O}$, and $\mathcal{N}$ is a normalization factor defined such that \eqref{statedef} has unit norm.

We now measure the energy at null infinity in this state.  In $d$ dimensions null infinity is a sphere $S^{d-2}$ and we parameterize it by a unit vector $n$.
\begin{equation}
\langle \mathcal{E}(n)\rangle_{\lambda \cdot \mathcal{O}} = \lim_{r\rightarrow \infty} r^{d-2}\int_{-\infty}^{\infty} dx^{-}~\langle \mathcal{O}(q,\lambda)|T_{--}(x^{-}, r n)|\mathcal{O}(q,\lambda)\rangle~. \label{Eintegral}
\end{equation}
The average null energy condition implies that the resulting function is non-negative as a function of the direction $n$.  

Since we are working in a conformal field theory this energy expectation value may be explicitly evaluated.  Indeed the object being integrated in \eqref{Eintegral} is a three-point function $\langle  \mathcal{O} T\mathcal{O}\rangle$ in Lorentzian signature with a prescribed operator ordering.  Thus, the result of \eqref{Eintegral} is an explicit function of OPE coefficients.

\subsubsection{External States Created by $T$}
\label{Tstates}

Let us review the essential details of this calculation in the case where the external state is created by an energy momentum tensor.  In general in $d\geq 4$ spacetime dimensions, the three-point function of energy-momentum tensors may be parameterized in terms of three independent coefficients 
\begin{equation}
\langle TTT \rangle =N_{B}\langle TTT \rangle_{B}+N_{F}\langle TTT \rangle_{F}+N_{V}\langle TTT \rangle_{V}~,
\end{equation}
where the various $B,F,V$ structures are those that arise in a theory of respectively free bosons, fermions, or $(d-2)/2$ forms.\footnote{In odd $d$ there is no free field associated to the structure parameterized by $N_{V},$ but nevertheless there is still a structure.  See \cite{Osborn:1993cr, Buchel:2009sk} for details.}  Our conventions are such that for free fields, $N_{B}$ counts the number of real scalars, $N_{F}$ the total number of fermionic degrees of freedom (e.g. it is $2^{\lfloor d/2\rfloor}$ for a Dirac fermion), and $N_{V}$ counts the number of degrees of freedom  in a $(d-2)/2$ form (for a single such field this number is $\Gamma(d-1)/\Gamma(d/2)^{2}$).   

A single linear combination of these coefficients is fixed by the conformal Ward identity, and related to the two-point function coefficient $C_{T}$ of energy momentum tensors (see equation \eqref{CTdef} for our conventions on the two-point function)
\begin{equation}
C_{T}=\frac{1}{\Omega_{d-1}^{2}}\left(\frac{d}{d-1}N_{B}+\frac{d}{2}N_{F}+\frac{d^{2}}{2}N_{V}\right)~.
\end{equation}
where $\Omega_n$ is the area of a sphere $S^n.$\footnote{ $\Omega_{n-1} = 2 \pi^{ n/2}/\Gamma(n/2)$.} As another point of reference let us briefly specialize to the case of four-dimensional theories.  In that case, the coefficients of the three-point function are related to conformal anomalies $a, c$ that parameterize the trace of the energy-momentum tensor in a general metric background
\begin{equation}
\langle T^{\mu}_{\mu}\rangle [g]=\frac{c}{16\pi^{2}}W^{2}-\frac{a}{16\pi^{2}}E^{2}~,
\end{equation}
where $W$ is the Weyl tensor and $E$ is the Euler density. The coefficient $c$ is proportional to $C_{T},$ while 
\begin{equation}
a= \frac{1}{1440}(4N_{B}+11N_{F}+124N_{V})~.
\end{equation}

Returning to case of general dimensions we now investigate the null energy operator using these three-point functions.  It is useful to organize the calculation using the relevant symmetries, which are rotations on the null $S^{d-2}$.  In addition, the three-point function of $T$'s is parity invariant.\footnote{In $d=3$ the three-point function has a parity odd piece which we discuss in section \ref{TTOd3}.}  It follows that the most general expression for the null energy is
\begin{equation}
\langle \mathcal{E}(n)\rangle_{\lambda \cdot T} =\frac{q}{\Omega_{d-2}}\left[1+t_{2}\left(\frac{\lambda^{*}_{ij}\lambda_{ik}n^{j}n^{k}}{|\lambda|^{2}}-\frac{1}{d-1}	\right)+t_{4}\left(\frac{\lambda^{*}_{ij}\lambda_{kl}n^{i}n^{j}n^{k}n^{l}}{|\lambda|^{2}}-\frac{2}{d^{2}-1}	\right)\right]~,
\end{equation}
where the constants have been fixed so that the total energy of the state is $q$, and $t_{2}$ and $t_{4}$ are computable functions of $N_{B}, N_{F}, N_{V}$.  

A useful way to understand the answer is to view the vector $n$ as fixed and to decompose the states (parameterized by their polarizations) under the remaining symmetry group $SO(d-2)$. 
For example, the polarization that has spin zero under rotations around the $\vec n$ axis is
\be
\lambda^{0}_{ij} \propto \left( n_i n_j - { \delta_{ij}  \over (d-1) } \right) \label{scalarpoldef}
\ee
In a similar way we can write polarization tensors that have spin one and spin two under rotation around the 
$\vec n$ axis. 
    The energy flux in the direction $n$ is the same for every state in a fixed $SO(d-2)$ 
    representation, and we denote them by $qT_{i}/\Omega_{d-2}$. Explicitly carrying out the integrals gives:
\bea
\label{Tsdef}
T_0 &=& \left(1-\frac{t_2}{d-1}-\frac{2t_4}{d^2-1}\right) +\frac{d-2}{d-1}(t_2 + t_4) = \rho_0(d) \left(\frac{N_B}{C_T}\right),\nn
T_1 &=& \left(1-\frac{t_2}{d-1}-\frac{2t_4}{d^2-1}\right) +\frac{t_2}{2} = \rho_1(d) \left(\frac{N_F}{C_T}\right),\\
 T_2 &=& 1-\frac{t_2}{d-1}-\frac{2t_4}{d^2-1}  = \rho_2(d) \left(\frac{N_V}{C_T}\right)~,\nonumber
 \ea
where the index labels the $SO(d-2)$ charge and in the above $\rho_{i}(d)$ is a positive function that depends only on the spacetime dimension (and not the OPE coefficients). Their explicit form is given in equation \eqref{appalphas}.

Additional symmetries imply constraints on the parameters above. In any superconformal field theory we have $t_4=0$. For holographic CFTs dual to Einstein gravity the parameters are $t_2=t_4=0$, giving angle independent energy one-point functions $T_0=T_1=T_2=1$.

Returning to the general discussion, we can see from \eqref{Tsdef} that the average null energy condition implies the inequalities
\begin{equation}
N_{B}\geq 0~, \hspace{.5in}N_{F}\geq 0~, \hspace{.5in}N_{V}\geq 0~. \label{Tineq}
\end{equation}

One significant remark concerning the bounds \eqref{Tineq} is that they may clearly be saturated in free field theories.  Conversely, it has been argued \cite{Zhiboedov:2013opa}  that any theory that saturates the conformal collider bounds must be free.   The fact that the bounds may be saturated in actual CFTs illustrates that the conformal collider is an efficient way of extracting the implications of the average null energy condition.  Namely, we could not possibly get a stronger bound, 
otherwise we would run into a contradiction with free theories.

\section{Bounds on $TT\mathcal{O}$ in $d\geq 4$}
\label{TTOdg4}

We now turn to our main generalization of the conformal collider bounds reviewed in section \ref{ccreview}.  We explore the consequences of the average null energy condition in more general states than those created by a single primary operator.  Specifically in this section we will investigate states which are obtained by a linear combination of primary operators.  We will find that the average null energy condition in such states yields new inequalities on OPE coefficients. 

In this section, the states we consider will be created by a linear combination of the energy-momentum tensor and a general scalar hermitian operator $\mathcal{O}.$ We parameterize such a state in terms of normalized coefficients $v_{i}$
\begin{equation}
| \Psi \rb = v_1 | T(q,\lambda) \rb + v_2 | \OO(q)\rb~.
\end{equation}
The energy one-point function in the collider experiment is now a matrix
\begin{equation}
\lb \Psi| \mathcal{E}(n) | \Psi \rb = v^\dagger \left(\begin{array}{cc} \lb T(q,\lambda) | \mathcal{E}(n) | T(q,\lambda)\rb & \lb T(q,\lambda) | \mathcal{E}(n) | \OO(q)\rb  \\ \lb T(q,\lambda) | \mathcal{E}(n) | \OO(q)\rb^* &\lb \OO(q) | \mathcal{E}(n) | \OO(q) \rb  \end{array}\right) v ~.\label{matrixenergy}
\end{equation}
The average null energy condition implies that this matrix is positive definite.  This is a stronger condition than requiring that the diagonal entries are positive and will imply new inequalities on OPE coefficients.

The majority of the entries in this matrix have already been computed.  For instance, in section \ref{Tstates} we reviewed the portion of the matrix involving the energy expectation value in states created by the energy momentum tensor.  Even simpler is the entry involving the expectation value in the scalar state which gives rise to a uniform energy distribution 
\begin{equation}
\lb \OO(q) | \mathcal{E}(n) | \OO(q)\rb=\frac{q}{\Omega_{d-2}}~.
\end{equation}

It remains to determine the off-diagonal entries in the matrix.  It is again useful to organize the expected answer using the rotation group on the null sphere.  Clearly we have
\begin{equation}
 \lb T(q,\lambda) | \mathcal{E}(n) | \OO(q)\rb\sim \lambda_{ij}n^{i}n^{j}~. \label{offdiag1}
\end{equation}
Therefore, the only polarization of the energy momentum tensor that participates in the non-trivial interference terms is the scalar $T_{0}$ aligned along the axis $n$ (see equation \eqref{scalarpoldef}).  

To extract this matrix element we require the three-point function $\langle TT\mathcal{O}\rangle$.  In all $d\geq 4$, the conservation constraints on $T$ imply that this correlator is fixed in terms of a single OPE coefficient $C_{TT\mathcal{O}}.$  We set conventions for our normalization of this OPE coefficient by examining a simple OPE channel.   Specifically we restrict all operators to a two-plane,  spanned by complex coordinates $z, \bar{z}$.  Then the OPE is
\begin{equation}\label{defOPETTO}
T_{zz}(z)T_{\bar{z}\bar{z}}(0)\sim \frac{C_{TT\mathcal{O}}}{|z|^{2d-\Delta}} \mathcal{O}(0)~.
\end{equation}
If we further assume that $\mathcal{O}$ is hermitian then the OPE coefficient $C_{TT\mathcal{O}}$ is real.  Additional details of this correlator including the full $d$-dimensional Lorentz covariant OPE and relation to the spinning correlator formalism of \cite{Costa:2011mg} are given in appendix \ref{app:TTO}. 

Based on these remarks, we can in general parameterize the energy flux in the direction $n$ coming from the off-diagonal matrix element \eqref{offdiag1} as
\begin{equation}
 \lb T(q,\lambda_0) | \mathcal{E}(n) | \OO(q)\rb= \frac{q}{\Omega_{d-2}}\left(\frac{C_{TT\mathcal{O}}}{\sqrt{C_{T}C_{\mathcal{O}}}}\spc h(\Delta)\right)~,
\end{equation}
where $h(\Delta)$ is some universal function that may be extracted from the conformal collider calculation, and the factors of $C_{T}$ and $C_{\mathcal{O}}$ arise from normalizing the states.  The relevant portion of the energy matrix \eqref{matrixenergy} is two-by-two and takes the form
\begin{equation}
\frac{q}{\Omega_{d-2}} \left(\begin{array}{cc} T_{0} &\frac{C_{TT\OO}}{\sqrt{C_T C_\OO}}\spc h(\Delta)  \\ \frac{C_{TT\OO}^*}{\sqrt{C_T C_\OO}} \spc h(\Delta) & 1 \end{array}\right) ~.
\end{equation}
Positivity of this matrix therefore leads to the constraint 
\begin{equation} \la{BoundFT}
\frac{|C_{TT\mathcal{O}}|^{2}}{C_{T}C_{\mathcal{O}}}\spc |h(\Delta)|^{2}\leq T_{0}~.
\end{equation}

More generally we may instead consider the collider experiment in a state created by $T$ plus a general linear combination of primary scalar operators.  Positivity of the resulting energy matrix is then equivalent to the following sum rule
\begin{equation}
\sum_{\text{Scalar Primaries}~\mathcal{O}_{i}} \hspace{-.1in}\frac{|C_{TT\mathcal{O}_{i}}|^{2}}{C_{T}C_{\mathcal{O}}}~ |h(\Delta_{i})|^{2}\leq T_{0}~.
\end{equation}
In appendix \ref{app:TTO} we explicitly compute the function $h(\Delta)$ (see equation \eqref{eq:defH}).  By combining the result with the expression \eqref{Tsdef}, we may reexpress the bound as 
\begin{equation}
\sum_{\text{Scalar Primaries}~\mathcal{O}_{i}} \hspace{-.1in}\frac{|C_{TT\mathcal{O}_{i}}|^{2}}{C_{\mathcal{O}}}~ f(\Delta_{i})\leq N_{B}~,\label{bound1}
\end{equation}
where $f(\Delta)$ is given as
\begin{equation} \la{Fvalue1}
f(\Delta)= \frac{(d-1)^3 d \pi ^{2 d} \Gamma \left(\frac{d}{2}\right) \Gamma (d+1) \Gamma (\Delta ) \Gamma \left(\Delta -\frac{d-2}{2}\right)}{(d-2)^2 \Gamma \left(\frac{\Delta }{2}+2\right)^4  \Gamma \left(\frac{d+\Delta }{2}\right)^2\Gamma \left(d-\frac{\Delta }{2}\right)^2}~.
\end{equation}

\subsection{Analysis of the Bound}
\la{AnBound}

We now turn to an analysis of the consequences of the general bound \eqref{bound1}.  The function $f(\Delta)$ has a number of significant properties.
\begin{itemize}
\item Expanded near the unitarity bound we find a first order pole:
\begin{equation}
f\left(\frac{d-2}{2}+x\right)\sim \frac{1}{x}~.
\end{equation}
Therefore in any family of theories, an operator $\mathcal{O}$ which is parametrically becoming free (i.e. $\Delta=(d-2)/2+x$ with $x$ tending to zero) must have $|C_{TT\mathcal{O}}|$ vanish at least as fast as $\sqrt{x}$.

\item For large $\Delta$ we find exponential growth
\begin{equation}
f(\Delta)\sim \frac{4^{\Delta}}{\Delta^{\frac{7d}{2}+4}}~.
\end{equation}
We may use this growth to approximate the sum in the bound for scalar operators of large $\Delta$.  Indeed, let $\rho(\Delta)$ denote the asymptotic density of scalar primary operators.  From convergence of the sum we then deduce that for large $\Delta$ the spectral weighted OPE coefficients must decay exponentially fast
\begin{equation}
\rho(\Delta)\frac{|C_{TT\mathcal{O}}|^{2}}{C_{\mathcal{O}}}\leq \frac{\Delta^{\frac{7d}{2}+3}}{4^{\Delta}}~.
\end{equation}
These estimates agree with those implied by convergence of the OPE expansion found in \cite{Pappadopulo:2012jk} for scalar operators.

\item  If $\Delta$ is an even integer greater than or equal to $2d$ we find that $f(\Delta)$ vanishes. We can understand the necessity of this as follows. We can imagine a large $N$ CFT dual to weakly coupled theory of gravity. In such theories we can consider the sequence of operators $ \mathcal{O} = : T^{AB} \partial^{ 2n } T_{AB}:$.  At large $N$ the dimensions of these operators are fixed to $\Delta = 2 d + 2 n$. Moreover, for these operators ${ C_{TT{\cal O} }^2 \over C_{\cal O }  } $ is of order $C_T^2$. Thus, compatibility with the bound \nref{bound1} for large $C_T$, requires that $f(\Delta)$ vanishes at these locations.   

The above argument does not explain why $f(\Delta)$ has double zeros. But the double zeros imply that the bound may be obeyed at subleading order, where we include the anomalous dimensions of these operators which scale as $1/C_T,$ by truncating the sum on $n$.\footnote{We thank E. Perlmutter for comments on this point.}

\item The function $f(\Delta)$ is non-zero for $\Delta=d$.  Therefore the bound \eqref{bound1} may be applied to marginal operators.  In that context, it constrains the change in $C_{T}$ at leading order in conformal perturbation theory.

\end{itemize}
\subsection{Free Field Theories and Destructive Interference}\label{sec:freefields}

Let us investigate the bound further in free field theories.  These examples are interesting because the bound \eqref{bound1} is saturated.  

Consider first a theory of a free real boson $\phi$ in  dimension $d$.  There is a $\mathbb{Z}_{2}$ global symmetry under which $\phi$ is odd and the energy-momentum tensor $T$ is even.  Therefore we need only consider scalars made from an even number of $\phi$'s.  Since the explicit expression for $T$ is quadratic in the free fields, the only possible scalars that may contribute to the bound are $:\phi^{2}:$ and $:\phi^{4}:$.  

By a simple inspection of the Wick contractions we deduce that $:\phi^{4}:$ has vanishing $TT\mathcal{O}$ correlation function\footnote{
The contractions imply that $\langle TT :\phi^4 : \rangle \propto \langle T :\phi^2: \rangle \langle T :\phi^2 :\rangle $, which is zero since two-point functions of
different operators vanish.}.  Meanwhile $:\phi^{2}:$ has 
\begin{equation}
\frac{|C_{TT\mathcal{O}}|^{2}}{C_{\mathcal{O}}}=\frac{(d-2)^{4}\Gamma(d/2+1)^{4}}{8\pi^{2d}(d-1)^{4}}~. \label{phisval}
\end{equation}
This exactly saturates the bound \eqref{bound1}.  

We can also consider the bound applied to free fields of different spin.  In $d=4$ the theory of free fermions or free gauge bosons have vanishing $N_{B}$.  Therefore the bound implies that for all scalar operators $\mathcal{O}$ either $C_{TT\mathcal{O}}$ vanishes, or $\mathcal{O}$ has dimension $2d+2n$ for non-negative integer $n$.  

It is straightforward to directly verify this prediction.  For instance consider the free vector.  The gauge invariant field strength gives rise to two local operators $F^{+}_{\mu\nu}$ and $F^{-}_{\mu\nu},$ which are respectively self-dual and anti-self-dual two-forms.  Note that this free field theory enjoys a continuous electromagnetic duality symmetry under which $F_{\mu\nu}^{\pm}$ rotate with opposite charge.  The energy-momentum tensor $T_{\mu\nu}$ is neutral under this transformation, and hence a scalar operator $\mathcal{O}$ with non-vanishing $C_{TT\mathcal{O}}$ must also be neutral.  If we recall that $F^+_{\mu\nu} F^{-\mu\nu} $ vanishes identically, then we see that the lowest dimension neutral scalar operator is $(F^{+}_{\mu\nu}F^{+\mu\nu})(F^{-}_{\alpha \beta}F^{-\alpha \beta})$.  Since this has dimension  eight, the weight function $f(\Delta)$ vanishes.  Moreover all other scalar operators that are neutral have larger even integer dimension.  Thus, the bound is obeyed.

A more physical way to understand why the bound  is saturated  in the free scalar theory is to visualize the state created by local operators.   
    
\begin{figure}[t!]
\begin{center}
\begin{tikzpicture}[scale=0.9,rotate=0]
\node at (0,0) {$+$};
\node at (0.5,-0.5) {$T_0$};
\draw[thick,->] (0,0) -- (1.3,1.3);
\draw[thick,->] (0,0) -- (-1.3,-1.3);
\node at (0,-2) {\small (a)};
\end{tikzpicture}
\hspace{1.4cm}
\begin{tikzpicture}[scale=0.9,rotate=0]
\node at (0,0) {$+$};
\node at (0.5,-0.5) {$T_1$};
\draw[thick,->] (0,0) -- (1.3,1.3);
\draw[thick,->]  (-0.5-0.2,-1-0.2) -- (-0.2,-0.5-0.2);
\node at (0,-1.2)  {\small $1/2$};
\draw[thick,->] (0,0) -- (-1.3,-1.3);
\draw[thick,->]  (0.5+0.2,0+0.2) -- (1+0.2,0.5+0.2);
\node at (1.2,-1.2 +1.2)  {\small $1/2$};
\node at (0,-2) {\small (b)};
\end{tikzpicture}
\hspace{1.4cm}
\begin{tikzpicture}[scale=0.9,rotate=0]
\node at (0,0) {$+$};
\node at (0.5,-0.5) {$T_2$};
\draw[thick,->] (0,0) -- (1.3,1.3);
\draw[thick,->]  (-0.5-0.2,-1-0.2) -- (-0.2,-0.5-0.2);
\node at (0,-1.2)  {\small $1$};
\draw[thick,->] (0,0) -- (-1.3,-1.3);
\draw[thick,->]  (0.5+0.2,0+0.2) -- (1+0.2,0.5+0.2);
\node at (1.2,-1.2 +1.2)  {\small $1$};
\node at (0,-2) {\small (c)};
\end{tikzpicture}\\
\vspace{1cm}
\begin{tikzpicture}[scale=0.9,rotate=0]
\node at (0,0) {$+$};
\node at (0.5,-0.5) {$\mathcal{O}$};
\draw[thick,->] (0,0) -- (1.3,1.3);
\draw[thick,->] (0,0) -- (-1.3,-1.3);
\node at (0,-2) {\small (d)};
\end{tikzpicture}
\hspace{1.4cm}
\begin{tikzpicture}[scale=0.9,rotate=0]
\node at (0,0) {$+$};
\node at (0.5,-0.5) {$J_1$};
\draw[thick,->] (0,0) -- (1.3,1.3);
\draw[thick,->]  (-0.5-0.2,-1-0.2) -- (-0.2,-0.5-0.2);
\node at (0,-1.2)  {\small $1/2$};
\draw[thick,->] (0,0) -- (-1.3,-1.3);
\draw[thick,->]  (0.5+0.2,0+0.2) -- (1+0.2,0.5+0.2);
\node at (1.2,-1.2 +1.2)  {\small $1/2$};
\node at (0,-2) {\small (e)};
\end{tikzpicture}
\hspace{1.4cm}
\begin{tikzpicture}[scale=0.9,rotate=0]
\node at (0,0) {$+$};
\node at (0.5,-0.5) {$J_0$};
\draw[thick,->] (0,0) -- (1.3,1.3);
\node at (1.5,0.9) {$\phi_1$};
\draw[thick,->] (0,0) -- (-1.3,-1.3);
\node at (1.5-2.2,0.9-2.2) {$\phi_2$};
\node at (0,-2) {\small (f)};
\end{tikzpicture}
\caption{ We consider operators with zero spatial momentum that create a pair of free particles. In (a,b,c) we consider a stress tensor operator. We examine 
the wavefunction along the direction specified by the long arrow and we decompose the stress tensor according to the spin around that axis. (a) The spin zero state
is obtained for scalars, spin one for fermions (b) and spin two for vectors or self-dual forms (c). (d) is the state produced by a scalar operator with can interfere
with (a). (e) is produced by 
a current with spin one along the observation axis and can interfere with (b). Finally (f) is a current with spin zero along the observation axis in a theory of scalars. It produces
two different real scalars in the back to back configuration and cannot interfere with (a).   }
\label{Pairs}
\end{center}
\end{figure}
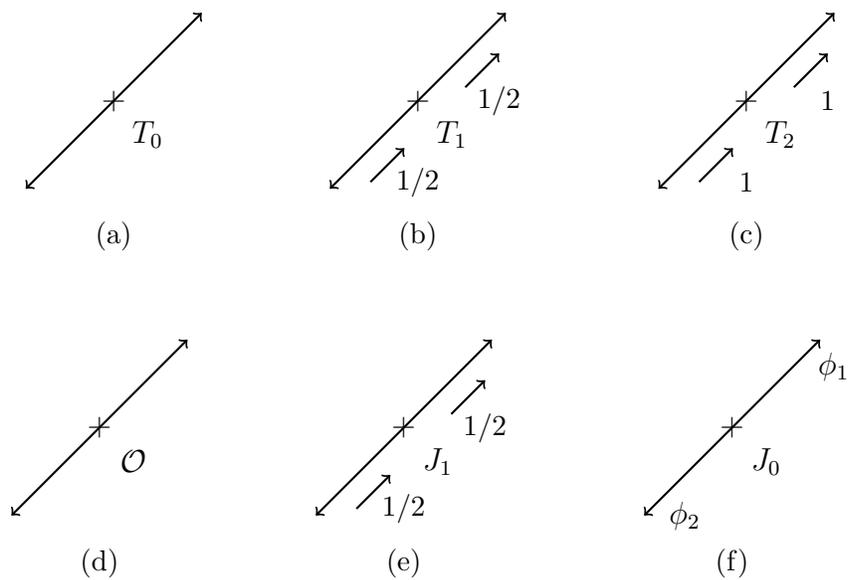

   Let us consider the action of an operator with non-zero energy but zero spatial momentum. If the operator is a bilinear in the fields, such as the stress tensor
   in a free theory, then it will create a pair of particles with back to back spatial momenta. Of course, the operator creates a quantum mechanical superposition of 
   states where these momenta point in various directions. For a scalar bilinear operator we get an s-wave superposition. For the stress tensor we get a superposition 
   determined by the polarization tensor. 
   
  As in previous sections, we measure the energy in the angular direction $n$ and hence can focus on the properties of the wavefunction for the pair of particles  in that particular direction. As in section \ref{Tstates} it is convenient to decompose the polarization tensors of the operators according to their angular momentum around the $n$ axis.    We can then easily check that a spin zero state $T_0$ can be produced only in a theory of scalars, a spin one state   $T_1$ can be produced only in a theory of fermions, and $T_2$ only in a theory of vectors (or $d/2-1$ forms),  see Figures \ref{Pairs}(a,b,c). This explains    formula  \ref{Tsdef}.

   A scalar operator of the form ${\cal O} = :\phi^2:$, where $\phi$ is an elementary scalar, can also produce a back to back combination of scalar particles, 
   see Figure \ref{Pairs}(d). Along the direction of
   observation this combination has the same form as the one produced by $T_0$, in Figure \ref{Pairs}(a). It is clear that we can make a quantum mechanical superposition 
   so that the wavefunction for the pair vanishes along that particular observation direction. This saturates the bound because 
     we get zero energy along that direction. For that   
   superposition of $T$ and ${\cal O}$ the energy along other directions is still non-zero.

   A similar argument helps us understand why we also saturate the $\langle TT J\rangle $ correlator 
   bound in the four dimensional theory of a Weyl fermion (see section \ref{TTJ}). 
   In that case we can make a superposition of the state $T_1$ in Figure \ref{Pairs}(b) with the state $J_1$ in \ref{Pairs}(e). Notice that we are using that $J$ couples 
   to a chiral fermion. If there was another fermion with the same helicity but opposite charge, as it would be the case for a vector-like current, then we would have an additional contribution to the state created by the current that will have a relative minus sign compared to the other charged particle pair. On the other hand, for the state created by the  stress tensor these two contributions have the same sign, therefore we cannot destructively interfere them. 
   
   This highlights that the bound comes from a quantum mechanical interference effect. We saturate the bound through a destructive interference effect that 
   prevents  particles from going into a particular direction. It is important to note that this is an interference for the pair of particles. For example, if 
   we consider a theory of scalars with a $U(1)$ symmetry generated by  a  current $J$, then in a basis of real scalars the current will create two different scalars, say 
   $\phi^1$ and $\phi^2$. This cannot interfere with the state created by the stress tensor where we have the same scalar for the two particles indicated in Figure
   \ref{Pairs}(a).

\section{Bounds on $TT\mathcal{O}$ in $d=3$}
\label{TTOd3}
In this section we will consider the case of $d=3$ separately. There are two reasons for doing this. First, the stress-tensor three-point function has two parity even structures, instead of three as in $d\geq4$, and has a parity odd piece which is special to $d=3$. Secondly, the correlation function $\lb TT\mathcal{O}\rb$ also has an extra parity odd structure special to $d=3$ \cite{Giombi:2011rz}.

First we consider external states created by the stress-tensor. We parametrize the three-point function of energy-momentum tensors as 
\beq
\lb TTT\rb = N_B \lb TTT\rb_B + N_F \lb TTT\rb_F + N_{\rm odd} \lb TTT\rb_{\rm odd}~, \label{odddef}
\eeq
where $N_B$ and $N_F$ already appeared in the $d\geq4$ case and $N_{\rm odd}$ parametrizes a new structure. We use the same convention for the explicit expression for $\lb TTT\rb_{\rm odd}$ as in 
\cite{Chowdhury:2017vel}\footnote{ We identify   our $N_{\rm odd}$ with their $\pi^4 p_T/3$.}.  In $d\geq4$ the energy one-point function of the collider experiment has a $SO(d-2)$ symmetry for the calorimeter direction $n$. The linearly independent tensor polarizations are organized as scalar, vectors or tensors with respect to this symmetry. In $d=3$ the group becomes $SO(1)$ and there are only two types of polarizations, which we take as 
\bea\label{eq:3dpolT}
\lambda_0 =\frac{1}{\sqrt{2}} \left(\begin{array}{cc}  1 & 0 \\  0 & -1 \end{array}\right)~,~~~~\lambda_1 &=&\frac{1}{\sqrt{2}} \left(\begin{array}{cc}   0 & 1 \\  1 & 0 \end{array}\right)~.  
\ea
The collider energy one-point function for an arbitrary polarization has the structure 
\beq
 \hspace{-0.412cm}\lb \mathcal{E}(n) \rb_{\lambda \cdot T} = \frac{q}{2\pi} \Bigg[1 + t_4 \left(\frac{|\lambda_{ij}n^in^j|^2}{|\lambda|^2}-\frac{1}{4} \right)+ d_4 \frac{\varepsilon^{ij}(n_i n^m \lambda_{jm}\lambda^*_{kp} n^k n^p + n_i n^m \lambda^*_{jm} \lambda_{kp} n^k n^p)}{2 |\lambda|^2}  \Bigg]~.
\eeq
To obtain a bound on these parameters we can consider a state created by $|\Psi \rb = v_1 | T(q,\lambda_0)\rb + v_2 | T(q,\lambda_1)\rb$. The energy matrix becomes  
\beq
\label{tmat3d}
\lb \Psi | \mathcal{E}(n) | \Psi \rb = \frac{q}{2\pi} v^\dagger \left(\begin{array}{cc} T_{0} & T_{\rm odd} \\ T_{\rm odd} & T_{1}  \end{array}\right) v~,
\eeq
where $T_{1} =1-t_4/4$, $T_{0} = 1+t_4/4$ and $T_{\rm odd}=d_4/4$. These parameters were computed in \cite{Chowdhury:2017vel} in terms of the $\lb TTT\rb$ parameters $N_B$, $N_F$ and $N_{\rm odd}$ obtaining 
\beq\label{3dTs}
C_T T_{1} = \frac{3}{16 \pi^2 } N_F~,~~ C_T T_{ 0} = \frac{3}{16 \pi^2 } N_B~,~~C_T T_{\rm odd} = \frac{3}{16\pi^2} N_{\rm odd}~. 
\eeq
 For supersymmetric CFTs $t_4=0$ just as in the case $d\geq 4$. Also, CFTs dual to Einstein gravity have $t_4=d_4=0$.

The average null energy condition implies that the matrix \eqref{tmat3d} is positive definite.  This implies $t_4$ and $d_4$ lie inside a circle $t_4^2 +d_4^2 \leq 4^2$, or equivalently $N_B\geq 0$, $N_F\geq 0$, and $N^2_{\rm odd} \leq  N_B N_F$.

Now we will generalize this construction along the same lines as presented in section \ref{TTOdg4}. We will consider a superposition between stress tensor and a scalar operator states 
\beq
|\Psi \rb = v_1 | T(q,\lambda_0)\rb + v_2 | T(q,\lambda_1)\rb + v_3 | \mathcal{O}(q)\rb~.
\eeq
As anticipated above, for $d=3$ the correlation function $\lb TT\mathcal{O}\rb$ is now determined by two parameters 
\beq
\lb TT\mathcal{O}\rb = C_{TTO}^{\rm even} \lb TT\mathcal{O}\rb_{\rm even} +C_{TTO}^{\rm odd} \lb TT\mathcal{O}\rb_{\rm odd}~,
\eeq
where the even part is given by specializing the arbitrary $d$ correlator $d=3,$ and our choice of normalization for the odd part is given explicitly in appendix \ref{app:d3odd}. We can make our conventions for this latter term as in \eqref{defOPETTO} in the following way. We can define $C_{TT\OO}^{\rm odd} $ by the following OPE
\be
T_{zz}(z, \bar z , y=0)  T_{z y}(0)  \sim  C_{TT\OO}^{\rm odd} { \bar z \over 4|z| }  \OO(0)~,
 \ee
 where the three spatial coordinates are $(z,\bar z, y)$. 
 
 Using this normalization, the energy one-point function is given in terms of a three-by-three matrix as 
\beq
\lb \Psi | \mathcal{E}(n) | \Psi \rb = \frac{q}{2\pi} v^\dagger \left(\begin{array}{ccc}  T_{0} & T_{\rm odd} &\frac{C^{\rm even}_{TTO}}{\sqrt{C_T C_{\mathcal{O}}} }h^{\rm even}_{3d}(\Delta) \\ 
T_{\rm odd} & T_{1} &  \frac{C^{\rm odd}_{TTO}}{\sqrt{C_T C_{\mathcal{O}}} } h^{\rm odd}_{3d}(\Delta)  \\
\frac{C^{\rm even}_{TTO}{}^*}{\sqrt{C_T C_{\mathcal{O}}} } h^{\rm even}_{3d}(\Delta)  & \frac{C^{\rm odd}_{TTO}{}^*}{\sqrt{C_T C_{\mathcal{O}}} } h^{\rm odd}_{3d}(\Delta)   & 1  \\  \end{array}\right) v~,
\eeq
where the functions $h^{\rm even}_{3d}(\Delta)$ and $h^{\rm odd}_{3d}(\Delta)$ can be obtained repeating the procedure reviewed in appendix \ref{app:TTO} and we obtain 
\bea
h^{\rm odd}_{3d}(\Delta) &=&\frac{12 \sqrt{6}  \pi^2\sqrt{\Gamma(2\Delta-1)}}{\Gamma(\frac{\Delta+1}{2})\Gamma(\Delta+3)} \frac{1}{\Gamma(\frac{7-\Delta}{2})} ~, \la{hodd}  \\ 
h^{\rm even}_{3d}(\Delta) &=&\frac{12 \sqrt{6}  \pi^{2}\sqrt{\Gamma(2\Delta-1)}}{\Gamma(2+\frac{\Delta}{2})\Gamma(\Delta+3)} \frac{1}{\Gamma(3-\frac{\Delta}{2})}~. \la{heven}
\ea
Demanding positive definiteness of the energy matrix gives several types of constraints which involve the scalar OPE coefficients. Two of these bounds are easy to generalize to an arbitrary number of scalar operators 
\beq\label{eq:3d2x2b}
\sum_i\frac{|C_{TT\mathcal{O}_i}^{\rm even}|^2}{C_{\mathcal{O}_i}} f_{\rm even}(\Delta_i) \leq  N_B ~,~~~~\sum_i \frac{|C_{TT\mathcal{O}_i}^{\rm odd}|^2}{C_{\mathcal{O}_i}} f_{\rm odd}(\Delta_i) \leq N_F~,
\eeq 
where we defined $f_{\rm odd/even} = | h_{3d}^{\rm odd/even}|^2 /3$. We can consider the positivity of the determinant of the $3\times 3$ matrix. This gives an independent bound which together with the bound on $\lb TTT\rb$ is sufficient for the positivity of the energy one-point function
\bea\label{eq:3d3x3b}
&& \hspace{-2cm}N_B \frac{|C_{TT\mathcal{O}_i}^{\rm even}|^2 f_{\rm even}(\Delta_i) }{C_TC_{\mathcal{O}_i}}+ N_F  \frac{|C_{TT\mathcal{O}_i}^{\rm odd}|^2f_{\rm odd}(\Delta_i)}{C_TC_{\mathcal{O}_i}}\nn
 &&  - 2 N_{\rm odd} \frac{{\rm Re}\spc[ C_{TT\mathcal{O}_i}^{\rm even}\sqrt{f_{\rm even}(\Delta_i)} C_{TT\mathcal{O}_i}^{\rm odd} \sqrt{f_{\rm odd}(\Delta_i)}]}{C_TC_{\mathcal{O}_i}}  \leq  N_BN_F - N_{\rm odd}^2\spc ~.
\ea
This bound can also be generalized to include an arbitrary number of scalar operators. However, as opposed to the situation in section \ref{TTOdg4}, the bounds involving different number of operators are independent. Their expressions in this case become more cumbersome and we will omit them here.

The \nref{hodd} \nref{heven} 
 have similar properties as the one appearing for the $d\geq4$ bound. Namely they diverge at the unitarity bound $\Delta =1/2$ and have zeros at $6+2n$ (even) and $7+2n$ (odd) for integer $n$. The zeros in the even case were explained by the existence of operators with two stress tensors in theories that are dual to weakly coupled gravity, see the last point in section \ref{AnBound}. The odd ones have the same explanation, except that now the scalar operators have the structure  $\epsilon^{ABC} T_{A D} \partial^{ 2 n} \partial_C T_{B D} $. 

\subsection{Chern-Simons Matter Theories}

In this section we apply the bounds derived to large $N$ Chern Simons theories at level $k$ coupled to fundamental matter. For definiteness we will consider fundamental fermions. We will denote the 't Hooft coupling by $\theta = \pi N/2 k$. The elements of the energy matrix involving the stress tensor were computed in \cite{Chowdhury:2017vel} using the explicit large $N$ expressions for the stress tensor three-point function \cite{Maldacena:2012sf}. The result is 
\beq
T_{1} = 2 \cos^2 \theta~,~~T_{0} = 2\sin^2 \theta~,~~T_{\rm odd} = 2 \sin\theta \cos\theta~.
\eeq
Using the conventions in, for example, \cite{Sezgin:2017jgm} we can compute the off-diagonal elements involving stress-tensor mixed with a scalar operator. In the fermionic theory we consider the scalar denoted by $\mathcal{O}\sim \psi \bar{\psi}$ has  dimension $\Delta =2$. The final result for the energy matrix is 
\beq\label{Ecsm}
\lb \Psi | \mathcal{E}(n) | \Psi \rb = \frac{q}{2\pi} v^\dagger \left(\begin{array}{ccc}  2 \cos^2\theta & 2 \sin \theta \cos \theta &\sqrt{2} \cos\theta \\ 
2\sin\theta \cos\theta & 2\sin^2\theta &  \sqrt{2} \sin \theta  \\
\sqrt{2} \cos \theta & \sqrt{2} \sin \theta  & 1  \\  \end{array}\right) v~.
\eeq
As a function of the 't Hooft coupling, this matrix has the property that all the minors have vanishing determinant. This implies saturation for all types of superposition of states. For the case of the stress tensor this was noted in \cite{Chowdhury:2017vel}, but we find that this is a more general feature for states where we also act with $\mathcal{O}$. 

Even though we do not have a concrete physical picture explaining this,  we expect a picture 
 along the lines of section \ref{sec:freefields}, where the interaction with the Chern-Simons gauge field has the effect of replacing free bosons or fermions by ``free anyons''. 

This discussion can also be applied to the case of CS coupled to fundamental bosons. From \cite{Maldacena:2012sf} we know that the energy matrix, given in terms of CFT three-point functions, can be obtained from the 
fermionic theory by the replacement $\theta \to \theta + \frac{\pi}{2}$ when  we consider the operator $\mathcal{O} \sim \phi^2$ of dimension $\Delta=1$. More generally we can consider the answer \eqref{Ecsm} as giving the energy matrix of a large $N$ theory with a slightly broken higher spin symmetry parametrized by $\theta$. 

\subsection{$3$d Ising Model}

\begin{figure}[t!]
\begin{center}
\begin{tikzpicture}[scale=0.85]
    \begin{axis}[
        xlabel=$C_{TT\varepsilon}/C_{\rm free}$,
        ylabel=$C_{TT\varepsilon'}/C_{\rm free}$,
        scale only axis,
        ymin=-15, ymax=15,
                xmin=-15, xmax=15,
        domain=-1.74967:1.74967
        ]
         \addplot[name path = A,smooth, thick]{(1-0.32665239*x*x)^(0.5)/(0.0062923035162)^(0.5)};
         \addplot[name path = B, smooth, thick]{-(1-0.32665239*x*x)^(0.5)/(0.0062923035162)^(0.5)};
    \addplot[only marks,black, mark size=0.23pt,mark=square] coordinates {(-1.74967,0)(1.74967,0)(1.74967,0.01)(1.74967,-0.01)};
   \addplot[blue!20] fill between[of=A and B];
    \end{axis}
    \end{tikzpicture}
    \vspace{0.5cm}
    \end{center}
    \vspace{-0.85cm}    
    \caption{ 3d Ising model allowed region for $C_{TT\varepsilon}$ and $C_{TT\varepsilon'}$. }
    \label{fig:ising}
    \end{figure}
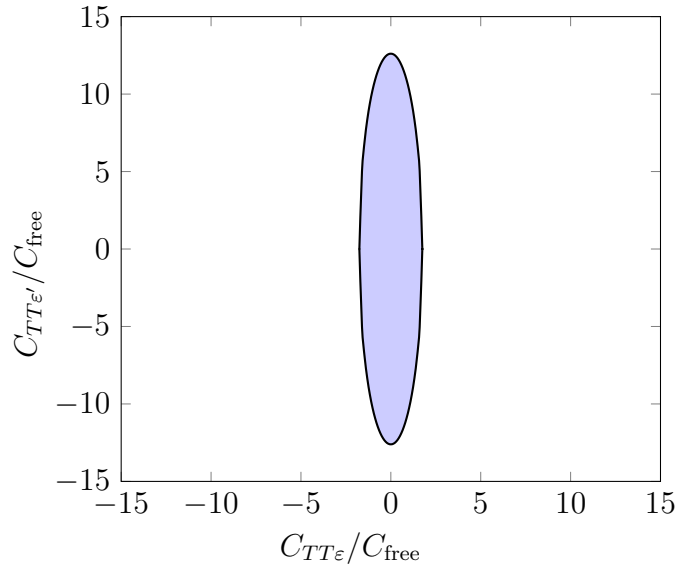
As another example, we can apply our bounds to obtain predictions for three-point coefficients for the $3d$ Ising model.  First let us parameterize the three-point coefficients of the energy-momentum tensor.  Since this theory is parity preserving the coefficient $N_{\text{odd}}$ in \eqref{odddef} is necessarily zero.  The remaining two structures in $\langle TTT\rangle$ have recently been computed numerically using the conformal bootstrap in \cite{El-Showk:2014dwa, Dymarsky:2017yzx}.  Explicitly\footnote{In making these estimates we use a value of $\theta \approx .014$.  This is the central value of the calculation of \cite{Dymarsky:2017yzx} based on expectations for the parity odd scalar gap.}
\begin{equation}
N_{B}\approx .9334~,\hspace{.5in}N_{F}\approx .0131~.
\end{equation}
The Ising model has a $\mathbb{Z}_{2}$ global symmetry under which $T$ is even.  Therefore only $\mathbb{Z}_{2}$ even scalars participate in the bound.  The lightest $\mathbb{Z}_{2}$ even and parity even scalar is the operator $\varepsilon$ whose dimension is known
\begin{equation}
\Delta_{\varepsilon}\approx 1.4127~.
\end{equation}
Therefore, in a normalization where the two-point function coefficient of $\varepsilon$ is one, we can evaluate \eqref{eq:3d2x2b} and find the bound
\begin{equation}
|C_{TT\varepsilon}|\leq .0088=(1.751) \spc C_{\rm free}~,
\end{equation}
where in the last equation we normalized the answer by the expression \eqref{phisval} for the value of the OPE coefficient in the free theory $C_{\rm free} =|C_{TT:\phi^{2}:}|/\sqrt{C_{:\phi^2:}}$ .  Note that although $:\phi^{2}:$ saturates the bound in the free field theory, the dimension of $\varepsilon$ is larger than that of $:\phi^{2}:$ and hence the OPE coefficient $C_{TT\varepsilon}$ may be larger than $C_{TT:\phi^{2}:}$. 
    
We can obtain a stronger bound by including the operator $\varepsilon'$ of dimension $\Delta_{\varepsilon'}\approx 3.8303$ in the sum of \eqref{eq:3d2x2b}. Using the correct values for $f_{\rm even}(\Delta)$ for these dimensions and normalizing by the $TT:\phi^2:$ OPE we obtain the constraint
\beq
0.3267|C_{TT\varepsilon}|^2   + 0.0063|C_{TT\varepsilon'}|^2   \leq C_{\rm free}^2~.
\eeq
Since the operators $\varepsilon$ and $\varepsilon'$ are hermitian their OPE coefficients are real and the bound above defines the allowed region of OPE coefficients as the interior of an ellipse shown in Figure \ref{fig:ising}.

\section{Bounds on $TTJ$ in $d=4$}
\label{TTJ}

As a final example, we consider states created by a linear combination of the energy-momentum tensor and a conserved vector current $J$  in $d=4$ spacetime dimensions.  In this case the three-point function $\langle TTJ\rangle$ is controlled by a single OPE coefficient $C_{TTJ}$ and is parity violating.  This three-point function is presented in detail in appendix \ref{TTJapp}.

One reason why this OPE coefficient is interesting is that it is equivalent to a non-trivial mixed anomaly between the flavor symmetry generated by $J$ and the Lorentz symmetry generated by $T$ \cite{Erdmenger:1999xx}.  In the presence of a background metric $g$, the current $J$ is not conserved but instead obeys \cite{Delbourgo:1972xb, Eguchi:1976db, AlvarezGaume:1983ig}
\begin{equation}
\langle \nabla^{\mu}J_{\mu}\rangle[g]=\frac{C_{TTJ}}{768\pi^{2}}\epsilon^{\mu\nu\rho\sigma}R_{\mu\nu\delta \gamma}R_{\rho\sigma}^{\phantom{\rho\sigma}\delta \gamma}~,
\end{equation}
where $R_{\mu\nu\rho\sigma}$ is the Reimann tensor.  

In the above, our normalization is such that the coefficient $C_{TTJ}$ may be expressed as the net chirality of the charges of elementary Weyl fermions:
\begin{equation}
C_{TTJ}=\sum_{\text {Left~Weyl~}i}q_{i}-\sum_{\text{Right~Weyl~}j}q_{j}~.
\end{equation}
In particular, for the theory of a single Weyl fermion $C_{TTJ}$ is one.  In an abstract CFT without a Lagrangian presentation our normalization of the OPE coefficient is defined as follows.  Fix complex coordinates $(z,w).$  Then the OPE of operators restricted to the $w=0$ plane is 
\begin{equation}
T_{ww}(z)T_{\bar{w}\bar{w}}(0)\sim \frac{C_{TTJ}}{4\pi^{6}|z|^{6}}\left(zJ^{\bar{z}}-\bar{z}J^{z}\right)~.
\end{equation}

We will also need the three-point function $\langle TJJ\rangle$.  This correlator is controlled by two independent coefficients:
\begin{equation}
\langle TJJ\rangle = Q_{CB}^{2} \langle TJJ\rangle_{CB}+Q_{WF}^{2} \langle TJJ\rangle_{WF}~.
\end{equation} 
Here the structures $CB$ and $WF$ are those found for the $U(1)$ current in a theory of free complex bosons ($CB$) or free Weyl fermions ($WF$).  In a free field theory, these are expressed in terms of the charges of elementary fields as (see \cite{Osborn:1993cr})
\begin{equation}
Q_{CB}^{2}=\sum_{\text{complex~scalars}~i}q_{i}^{2}~, \hspace{.5in}Q_{WF}^{2}=\sum_{\text{Weyl fermions}~i}q_{i}^{2}~.
\end{equation}
In general, a single linear combination of these OPE coefficients is fixed by the Ward identity.  We have
\begin{equation}
\langle JJ\rangle\propto C_{J}\equiv \frac{1}{3}\left(Q_{CB}^{2}+2Q_{WF}^{2}\right)~.
\end{equation}
The two-point function coefficient $C_{J}$ can also be interpreted as a conformal anomaly.  Indeed, in the presence of a non-trivial background gauge field that couples to $J$, the energy-momentum tensor acquires an anomalous trace.  In our conventions this is
\begin{equation}
\langle T_{\mu}^{\mu}\rangle[A]=\frac{C_{J}}{4}F^{\alpha\beta}F_{\alpha \beta}~. 
\end{equation}

We can bound the anomaly coefficient $C_{TTJ}$ using the same methods described in earlier sections for scalar operators.  We enforce positivity of the average null energy operator $\mathcal{E}$ in the state $| \Psi \rangle$ created by a linear combination of $T$ and $J$
\begin{equation}
| \Psi \rangle =  | T(q,\lambda_T) \rangle +  | J(q,\lambda_J)\rangle.
 \end{equation}
The expectation values $\langle \mathcal{E}\rangle_{\lambda_{T}\cdot T}$ and $\langle \mathcal{E}\rangle_{\lambda_{J}\cdot J}$ have been computed in \cite{Hofman:2008ar}.  The matrix of energy expectation values in the states $| \Psi \rangle$ may again be decomposed in terms of the $SO(2)$ rotation symmetry about the vector $n$.  The current operator $J$ contributes states of charge $-1, 0, 1.$  As in the review of section \ref{Tstates} we may express the null energy expectation value as $(qJ_{i}/4\pi)$ where $i$ is the $SO(2)$ charge.  One then finds
\begin{equation}
J_{\pm 1} = \frac{Q_{WF}^{2}}{C_{J}}~.
\end{equation}

By repeating the collider calculation we find that the new off-diagonal matrix element is given by  
\begin{equation}
\label{tjoffdiag}
\langle T(q,\lambda_T)| \mathcal{E}(n) |J(q,\lambda_J) \rangle =\frac{q}{4\pi}\left( \sqrt{\frac{5}{\pi^4}}\frac{C_{TTJ}}{\sqrt{C_T C_{J}}}  \varepsilon_{ijk} \lambda^*_{T,im} \lambda_{J,k} n^m n^j\right)~.
\end{equation}
 Note that this structure is parity odd as expected.  There are other allowed parity odd expressions in terms of
 $\lambda_{ij}$ and $n^i$, but they do not arise in the null-energy matrix element.  An important feature of \eqref{tjoffdiag} is that only those states of $SO(2)$ charge $\pm1 $ can mix with the energy-momentum tensor.  In particular, 
  this means that bound will only involve the coefficient $T_{1}$ defined in \eqref{Tsdef}.  

Explicitly choosing appropriate polarization tensors we then find that positivity of the null energy matrix $\mathcal{E}$ leads to a single constraint on these OPE coefficients: 
\begin{equation}
C_{TTJ}^{2}\leq Q_{WF}^{2}N_{WF}~,
\end{equation}
where $N_{WF}=N_{F}/2$ counts the effective number of Weyl Fermions in the $\langle TTT\rangle$ correlation function.  This bound is saturated in the free field theory of Weyl fermions.  This can be understood using the interference argument described in section \ref{sec:freefields}.
 
\subsection{Supersymmetry and the $R$-Current}

As in our analysis of scalar operators, we can generalize these results to states created by multiple currents. This is particularly interesting in the case of supersymmetric theories.

In supersymmetric theories, there is always a current $J_{R}$ contained in the same supermultiplet as $T$. 
 In particular, since it resides in a different multiplet it can be distinguished from an ordinary flavor current $J_{F}$.  We would like to improve our bound on the trace anomaly of $J_{F}$ to account for the fact that the $R$-current $J_{R}$ always exists.  In order to do this we consider the state created by 
\beq
| \Psi \rb = v_1 |T(q,\lambda_T) \rb + v_2  | J_R(q,\lambda_J)\rb + v_3 | J_F(q,\lambda_J)\rb~.
\eeq
The new ingredient appearing in the calculation of the energy matrix corresponding to this state involves the three-point function $\lb T J_R J_F\rb$.    Using superconformal invariance we can fix this correlator completely. Since the details are not very illuminating we will outline the procedure. The number of parity even structures, two of them, coincides with the ones appearing in $\lb T JJ\rb$, namely relaxing permutation symmetry does not add new structures \cite{Costa:2011mg}. Moreover, using supersymmetric Ward identities \cite{Osborn:1998qu} one can check that no parity odd structure is allowed for $\lb T J_R J_F\rb.$\footnote{This is not true for a  three-point function
 of a stress tensor and two different conserved currents $\lb T J_1 J_2\rb$ in a generic theory. } Out of the two OPE coefficients characterizing $\lb T J_R J_F\rb$, a linear combination of them is related to the two-point function $\lb J_R J_F\rb$, which vanishes due to superconformal invariance. This leaves $\lb T  J_R J_F\rb$ fixed by a single OPE coefficient. Finally, since $J_R$ lies in the same multiplet as the stress tensor we can relate this number to $C_{TTF}$, the mixed anomaly generated by the flavor current.

Combining the results outlined in the previous paragraph, and the fact that there is no new structure  involved in the collider calculation, it is straightforward to obtain the off-diagonal matrix element 
\beq
\lb J_R (q,\lambda_J) | \mathcal{E}| J_F(q,\lambda_J)\rb = \frac{q}{4\pi} \left(\sqrt{\frac{20}{3 \pi^4}} \frac{C_{TTF}}{\sqrt{C_T C_F}} \right),
\eeq
 where we chose $n=(1,0,0)$ and $\lambda_J = (0,1,i)$ for definiteness. 
 
 We can express parameters related to the $R$-current in terms of $a$ and $c=C_T \pi^4/40$. The two-point function is related to $C_T$ by a supersymmetry Ward identity as $C_R = \frac{16}{3} c$. Its mixed anomaly is also fixed by supersymmetry to $C_{TTR} = 16 (c-a)$. Finally the energy one-point function is given by $J^R_{\pm 1} = \frac{a}{c}$ \cite{Hofman:2008ar, Hofman:2016awc}. Supersymmetry also fixes this parameter for flavor currents as $J^F_{\pm 1} =1$. Taking these facts into account allows us to write down the energy matrix as a function only of $a$, $c$, $C_{TTF}$ and $C_F$. We obtain
 \beq\label{susymatrix}
\lb \Psi | \mathcal{E}| \Psi \rb =\frac{q}{4\pi}v^\dagger \left(\begin{array}{cccc} \frac{2c-a}{c} & \sqrt{3} \frac{c-a}{c} & \frac{1}{\sqrt{2c}}\frac{C_{TTF}}{\sqrt{C_{F}}}   \\  \sqrt{3} \frac{c-a}{c} & \frac{a}{c} & \frac{1}{\sqrt{6c}}\frac{C_{TTF}}{\sqrt{C_{F}}}\\  \frac{1}{\sqrt{2c}}\frac{C_{TTF}}{\sqrt{C_{F}}}  & \frac{1}{\sqrt{6c}}\frac{C_{TTF}}{\sqrt{C_{F}}} & 1 \end{array}\right)v~,
\eeq
where for definiteness we have chosen $\lambda_J=(0,1,i)$ and a tensor polarization with the same $SO(2)$ spin. 

Enforcing the positivity of this matrix yields several constraints.  The leading two-by-two minor involving states $|T(q,\lambda_T)\rb + | J_R(q,\lambda_J) \rb$ gives the bound 
 \beq
 \frac{1}{2} \leq \frac{a}{c} \leq \frac{3}{2}~,
 \eeq
 which coincides with those derived in \cite{Hofman:2008ar}. This bound is saturated by a free chiral multiplet, $ \frac{a}{c}=\frac{1}{2},$ or a free vector multiplet, $ \frac{a}{c}=\frac{3}{2}$. 
 
 To constrain the gravitational anomaly coefficient we evaluate the determinant of the full three-by-three matrix \eqref{susymatrix}. This gives the following bound on the mixed anomaly for a flavor current
 \beq
 \left( \frac{a}{c}-\frac{1}{2}\right) \left( 36c-24a-\frac{C_{TTF}^2}{C_F}\right) \geq 0~.
 \eeq 
 For a free chiral multiplet the bound is automatically saturated, since the first term in the left hand side vanishes independently of $C_{TTF}$. Therefore we will assume that $\frac{a}{c} >\frac{1}{2}$. Then we obtain the following bound 
 \beq
 \frac{C_{TTF}^2}{12 \spc C_F} \leq 3c-2a~,
 \eeq
 which is stronger than the one derived in the previous section, without the use of supersymmetry. Note also that this is consistent with the free vector multiplet.  In that case the right-hand-side vanishes, but there are also no flavor currents.
 
To conclude this section, we can mention some contexts where such bound on the mixed anomaly is relevant. First of all, when we consider holographic CFT this anomaly is related to a 5d Chern-Simons term of the form $\int A \wedge R \wedge R$, where $A$ is the gauge field dual to the current $J$ (we will see in the next section how our bounds translate to bounds on the gravity couplings for the case of $TT\mathcal{O}$). 
 
 Finally, in the context of hydrodynamics and transport, quantum anomalies induce a special type of transport coefficients, see \cite{Son:2009tf} and, in particular, for the mixed anomaly \cite{Vilenkin:1978is,Landsteiner:2011cp,Landsteiner:2017lwm}. The coefficient bounded in this section $C_{TTJ}$, 
 is related to the mixed anomaly  recently observed experimentally in Weyl semimetals \cite{Gooth}. In the linear response regime, the mixed anomaly produces an energy current $\vec{j}$ given by \cite{Vilenkin:1978is,Landsteiner:2011cp,Landsteiner:2017lwm}
 \beq
 \vec{j} = 24 C_{TTJ}  T^2 \vec{B}~,
 \eeq
 where we denote the temperature by $T$ and the system is placed in a fixed magnetic field $\vec{B}$. This allows us to translate our results into concrete bounds for transport coefficients.

\section{Bounds on Coefficients of the $AdS$ Effective Action } 
\la{AdSsection}

If the $d$ dimensional boundary theory has an $AdS_{d+1}$ dual, then we would like to translate the bounds on $C_{TT{\cal O } }$ to 
bounds   on the coefficients of the bulk effective action. 
We are imagining that the theory has a large $N$ expansion. Then, to leading order, 
   the bulk is given by a collection of free fields propagating on the $AdS$ metric. 
The simplest interactions correspond to bulk three-point interactions. These lead to three-point functions in the boundary theory.
 For the case of gravitons we have a three-point interaction coming from the Einstein Lagrangian, but it is also  necessary to include higher derivative terms, of
the form $W^2$ and $W^3$,  in order
to get the most general structures for the tensor three-point function.    
It is possible to match the coefficients of the new structures to the coefficients of these higher derivative terms in the Lagrangian \cite{Hofman:2008ar,Buchel:2009sk}.

 Here we consider the same problem for the case of the $\langle TT {\cal O} \rangle $ correlator. The first observation is that in Einstein gravity this
 correlator is  zero, since the action of  any field, expanded around the minimum of its potential  has an action without a linear term in the scalar field. 
 Notice that  this also implies that a massive scalar field cannot not decay into two gravitons. However, we can   write the higher derivative term
 \be
 \la{ExtraT} 
 S= M_{pl}^{d-1} \alpha \int d^{d+1}x \sqrt{g}  \chi W^2
 \ee
  in the action, where we normalized the $\chi$  field to be dimensionless.\footnote{
Here $M_{pl}$ is the reduced Planck mass in $d+1$ dimensions, defined so that the Einstein term is $S = { M_{pl}^{d-1} \over 2} \int d^{d+1} x \sqrt{g} R $.
Similarly, the action of the scalar field is $ S =  { M_{pl}^{d-1} \over 2 }  \int  [( \nabla \chi)^2 - m^2 \chi^2 ]$. }
  This term enables the field $\chi$ to decay into two gravitons.
  In flat space there is only one structure for the on shell three-point function between a scalar and two gravitons, except in four dimensions where threre is also 
  a parity odd one, as we discuss later. Therefore the vertex \nref{ExtraT} represents the general interaction that we can have in the theory. There can be other
  ways to write it which give the same three-point function as  \nref{ExtraT}.  
   It is possible to check that \nref{ExtraT} gives rise to a  $\langle TT{\cal O } \rangle $ three-point function with the coefficient 
 \be \la{boundco} 
  { C_{TT {\cal O}} \sqrt{f (\Delta )}  \over 
 \sqrt{C_{T} } } = \frac{8\spc \sqrt{2d}(d-1)\pi^{d/2}}{\sqrt{d+1} \spc \Gamma(d/2)} { \alpha \over R^2_{AdS}    }   ~.
  \ee
 At first sight, it seems 
  surprising that the function  $f(\Delta) $ appearing here is the same as the one that appears in the bound \eqref{bound1}. 
  This means that the $\Delta$ dependence disappears when we express the bound in terms of $\alpha$. 
 This is easy to understand when we derive \nref{boundco} as follows. 
 
  First we notice that  integrating  the stress tensor along a null line, as in the definition of the energy measurement ${\cal E} = \int dx^- T_{--}(x^-,x^+=0,\vec y=0)$,
    we produce a shock wave in the bulk  that is localized at $x^+=0$. 
   We can then imagine scattering a superposition of $\chi$ and a graviton through this shock wave. This  leads to a time delay that is given by a 
   matrix mixing the graviton and the scalar. An important point is that the propagation through the shock wave is given by 
   integrating the wave equation in a small interval before and after $x^+=0$. Only the shock wave contributes to this short integral over $x^+$, but the scalar 
   mass term does not contribute. Therefore the time delay matrix is independent of the mass of the scalar. We can determine the precise coefficient in \nref{boundco} by doing this explicit computation for Einstein gravity  plus \nref{ExtraT}. 
   We then get a bound on $\alpha$ by requiring that the time delay is positive. Comparing this to the bound \eqref{bound1} we fix the coefficient to the one in 
   \nref{boundco}. We explain this in more detail in appendix \ref{GravBound}. 
   
   This same shock wave method enables one to set even stricter bounds on $\alpha$ if one assumes that there is a gap to the higher spin particles,\footnote{We thank E. Perlmutter and D. Meltzer for discussions on this issue.} as was discussed in   \cite{Camanho:2014apa} for the  case of the graviton higher derivative interactions. A similar analysis can be done for the 5d Chern-Simons term coupling dual to the mixed anomaly \cite{Bhattacharyya:2016knk}.

  In string theory, we expect that $\alpha $ is  the order of $\alpha'$, the inverse string tension. 
If gravity is a good approximation, $\alpha' \ll R^2$, 
 then we find that the bound on \nref{boundco} is far from being saturated.
 The bound is saturated only as the string length becomes of the order of the radius of $AdS$. 
 In particular, this implies that the bound is satisfied,  and far from being saturated, for the Konishi operator of ${\cal N}=4$ super Yang Mills at 
 strong coupling.  This operator is the lightest non-protected single trace operator which has a dimension growing like $\Delta \propto \lambda^{1/4}$ at strong coupling, 
 $\lambda \gg 1$. 
  
  In the four dimensional case, we can also have a parity odd correlator with a corresponding coupling.
  In flat space this is related to the fact that    the three-point functions with $++$ or $--$ graviton helicities are Lorentz invariant by themselves. (The $-+$ graviton
  helicities are forbidden by angular momentum conservation). 
 We can then  write the action as 
        \be \la{GravFd}
               S =   M_{pl}^2 \int d^4 x \sqrt{g} \left[ { 1 \over 2 }   (R -2 \Lambda)  + { 1 \over 2 }   [ (\nabla \chi)^2  -  m^2 \chi^2]
                 +   \int \alpha_e \chi W^2 + \alpha_o \chi  W W^* \right]~,
               \ee
  where as above we have defined $\chi$ to be dimensionless.\footnote{ We also
  define $(W^*)_{\mu \nu \rho \sigma } = \half \epsilon_{\mu \nu \delta \gamma } W^{\delta \gamma}_{~~\rho \sigma } $. }
   In this normalization $\alpha_i$ has dimensions of length squared. 
  They can be related to the coefficients of the three-point function as 
      \be \la{fdCoef}
      { C_{TT {\cal O}}^{\rm even} h_{\rm even}(\Delta) \over \sqrt{C_{T} } } = \frac{24}{\sqrt{2}}{ \alpha_e  \over   R^2_{AdS_4}} ~,~~~~~  {C _{TT {\cal O}}^{\rm odd} h_{\rm odd}(\Delta) \over
      \sqrt{ C_{T} } } 
      = \frac{24}{\sqrt{2}} { \alpha_o  \over  R^2_{AdS_4}}~.
         \ee
   The bounds in this case then read 
  \be \la{AdSBound}
 { \sqrt{   \alpha_e^2 + \alpha_o^2 } \over R_{AdS_4}^2}   \leq {1 \over 12 \sqrt{2}}  ~,
 \ee
 in the case that there are no purely gravitational  corrections to Einstein gravity. Of course, if there are three-point functions that lead to corrections to Einstein gravity, then 
 the bound is corrected to those given in section \ref{TTOd3}.
  
 \section{ Constraints for de-Sitter and Inflation } 
 \la{dSsection}
 
 The physics of inflation might be our very best window into very high energy physics. 
 The standard inflationary theory starts with a scalar field coupled to the Einstein action and includes all two (or less) derivative interactions.
 The universe undergoes a period of expansion that is governed by a nearly de-Sitter solution, characterized by a Hubble scale $H$ that is nearly constant. 
 The effective coupling of the gravitational sector is of order $H/M_{pl}$ which is very small, less than $10^{-5}$. 
 However, it is possible that there are corrections to the two derivative action due to the presence of a light string scale. The value of the string tension could be 
 fairly low $H^2 \lesssim T $. When the string tension becomes comparable to the Hubble scale, we   expect significant corrections to the two derivative action. 
 We do not have an explicit scenario where this happens. However, a similar situation happens in $AdS$ space when we consider a gravity dual of a not so strongly 
 coupled large $N$ theory. Therefore it is natural to question whether something similar could happen in inflation and we can look for signatures of such a 
 low string scale. 
 It is important to find signatures that are as model independent as possible. Specially nice signatures are those that have a non-vanishing contribution in the 
 de-Sitter approximation. These are not strongly suppressed by slow roll factors. In addition, their form is strongly constrained by the de-Sitter isometries. 
 An example of such contributions are the three-point functions of gravity fluctuations, where the higher derivative corrections were discussed in \cite{Maldacena:2011nz}. 
 Another interesting case are the couplings of the form $ f_e(\chi) W^2$ or  $f_0(\chi) W W^*$. 
  These two couplings are particularly interesting because their effects are visible at the two-point function level. 
 
 Let us discuss first the parity odd coupling, which 
   leads to chiral gravity waves \cite{Lue:1998mq,Alexander:2004wk}. Namely, we have  different gravity wave 
 two-point functions, $hh_{L}$, $hh_R$,  for the left and
 right handed circularly polarization. We can define the asymmetry $A$  as 
  \be  \la{AsymGW}
   A \equiv {hh^L - hh^R \over hh^L + hh^R}   =  4 \pi { \dot f_o(\chi) \over H }  {H^2  } = \pm 4 \pi \sqrt{ 2 \epsilon} \left( {\partial f \over \partial \chi } \right) H^2 ~,~~~~~~~~ \chi = { \phi \over M_{pl}} ~,
   \ee 
  where $\chi$ is defined to be dimensionless and $\phi$ is the inflaton with canonical normalization. (The $\pm$ comes from going from $\dot \chi$ to $\sqrt{\epsilon}$, 
  since  the  derivative of the scalar can have either sign). 
   If we were in $AdS_4$ we would have a sharp bound on the coefficients via the condition \eqref{AdSBound}, after we identify  $\alpha_o ={\partial f \over \partial \chi } $.
    It is reasonable to think that in the de-Sitter case too, there will be trouble is the bound is violated. Of course, we know that even near-saturation of the bound 
    implies that the field theory approximation is breaking down. 
    
     In the de-Sitter case we do not have a sharp derivation of
   a bound from boundary theory  reasoning. We do not have an analog of the null energy condition, discussed in section \ref{TTOd3}, for the boundary theory, since the 
   boundary theory is purely spacelike. It would be nice to have a sharp derivation of a de-Sitter version of  the bound. 
   In de-Sitter, we can talk of a ``quasi-bound'', which we get by simply applying the same bound on the coefficients 
   of the action that we had in anti-de-Sitter. This quasi-bound should be viewed simply as  an educated guess, including numerical coefficients, for the validity of bulk effective theory. 
   A near saturation of these quasi-bounds is a strong indication of a light string scale which could also have other manifestations such as indirect evidence of 
   higher spin massive particles, etc \cite{Arkani-Hamed:2015bza}.  
   In summary, 
    in de-Sitter also we have a quasi-bound on the coefficients similar to \nref{AdSBound}, with $1/{R_{AdS}} \to H$
   \be
   \la{dSBound}
 \sqrt{  \left( { \partial f_e \over \partial \chi } \right)^2 + \left( { \partial f_o \over \partial \chi } \right)^2  } = 
  \sqrt{   \alpha_e^2 + \alpha_o^2 } \leq {  H^2 \over  12 \sqrt{2} } ~.
 \ee
   This bound, then implies a quasi-bound on the asymmetry \nref{AsymGW}  of the form 
   \be
   | A|  \leq  { 4 \pi \over 12 } \sqrt{ \epsilon} ~.
     \ee
   The allowed values by this quasi-bound seem to be smaller than the smallest possible  measurable value 
    from the CMB B-modes as analyzed in \cite{Saito:2007kt}. 
    Conversely, this means that if chiral gravity waves  through E-B mode correlators are measured, then we
    would need a higher derivative coupling with a coefficient so large that it violates \nref{dSBound}.

     Let us turn now to a discussion of the parity even coupling. This coupling 
       gives rise to a violation of the consistency condition for the two-point function  \cite{Baumann:2015xxa}, even in the case
  that the speed of sound is close to one,   
  \be \la{Corr}
 - 8   { n_t \over r }   = 1 + 8 H^2 { H d_t f_e \over (  d_t \chi )^2  } = 1 \pm  8 H^2 { \alpha_e  \over \sqrt{2 \epsilon} } ~,
  \ee     
 where we assumed that the speed of sound for the scalar is close to one. Here $n_t$ is the tensor spectral index and $r$ the tensor to scalar ratio
 conventionally defined. 
  Then the bound we had in \nref{dSBound} translates into the following constraint on the violation of the consistency condition
  \be
        \left|- 8   { n_t \over r }   -1 \right| \leq { 1     \over 3 \sqrt{   \epsilon } } ~.
   \ee

   \subsection{Comments on Scalar-Tensor-Tensor Three-Point Functions } 
    
   The  $\phi W^2$ higher derivative coupling  between the scalar and the graviton also 
     give rise to new contributions to the scalar-tensor-tensor three-point function. 
  This is a contribution, that is non-vanishing in the de-Sitter limit. More precisely, if we can    approximate $\partial_\chi f_e(\chi(t))$ by a constant, then 
  we get a contribution even in de-Sitter space. 
  The standard Einstein gravity contribution, \cite{Maldacena:2002vr}, is suppressed by a slow roll factor $\sqrt{\epsilon}$, if we assume that $\partial_\chi f $ is of order one. 
  Of course, our bound constrains the size of this three-point function because it is constraining the size of the coefficient $\alpha_e \sim \partial_\chi f_e(\chi(t))$. 
  
  The three-point function for the 
  parity odd  coupling $f_o(\chi)  W W^*$ was computed in \cite{Bartolo:2017szm}, where it was found to be proportional to 
  $\partial_\chi^2 f$. One might have naively expected a de-Sitter invariant contribution proportional to $\alpha_0 = \partial_\chi f_0$, when we approximate this by a constant. The explicit computation by \cite{Bartolo:2017szm} shows that there is no such contribution. This seems surprising at first sight because this 
  parity odd coupling does indeed give a non-vanishing contribution to the three-point function in the $AdS_4$ case. The reason it vanishes in de-Sitter is that 
  it gives a contribution to the de-Sitter wavefunction that is a pure phase, which disappears when we take the absolute value squared of the wavefunction.  
  The same happens with the $W^2 W^*$ parity violating graviton three-point coupling \cite{Soda:2011am}.
   The correlator proportional to $\partial_\chi^2 f$ found in \cite{Bartolo:2017szm} has an extra factor of  $\dot \phi$ and is not expected to be de-Sitter invariant
   (though we did not check this explicitly). 
          
  It should be noted that the correction to the two-point function  consistency condition  \nref{Corr}  has the right form so that the 
  consistency condition involving the soft limit of the three-point function \cite{Maldacena:2002vr,Creminelli:2004yq} is obeyed, though we have not explicitly 
  checked the precise numerical coefficients. A similar remark applies in the parity odd case; the correction to the two-point function \nref{AsymGW} is 
  such that the soft limit of the three-point function in \cite{Bartolo:2017szm} obeys the consistency condition.

 \section*{Acknowledgements} 

We thank H. Casini, S. Giombi, R. Meyer, E. Perlmutter, D. Simmons-Duffin, and D. Stanford for discussions. We also thank E. Perlmutter for comments on a draft. C.C. is supported by the Marvin L. Goldberger Membership at the Institute for Advanced Study, and DOE grant DE-SC0009988. J.M. is supported in part by U.S. Department of Energy grant DE-SC0009988 and the Simons Foundation grant 385600.

\begin{appendix}

\section{Absence of Positive Local Operators}
\label{secnoloc}

Let us review the essential steps of \cite{Epstein:1965zza} in a modern language.  Let $\Phi$ be any Hermitian operator and $|0\rangle$ the Lorentz invariant vacuum state.  We make two assumptions:
\begin{itemize}
\item The one-point function $\langle 0| \Phi |0\rangle$ vanishes.
\item For all states $|\psi\rangle$ in the Hilbert space, the expectation value $\langle \psi|\Phi|\psi\rangle$ is non-negative.
\end{itemize}
Under these assumptions we may prove that $\Phi$ annihilates the vacuum state, $\Phi|0\rangle=0.$  Indeed, for any positive operator, the Cauchy-Schwarz inequality implies that
\begin{equation}
|\langle \psi|\Phi|0\rangle|^{2}\leq \langle \psi|\Phi|\psi\rangle\langle 0|\Phi|0\rangle~.
\end{equation}
Since the right-hand side is zero by hypothesis, we conclude that $\Phi|0\rangle$ must vanish.

If we now further assume that $\Phi$ is an operator localized within a compact region ${\cal R}$, we can deduce that $\Phi$ must vanish. 
 To demonstrate this, consider operators localized in  a 
region ${\cal R}'$ that is spacelike separated from ${\cal R}$.  Let us denote by $O_{{\cal R}'}$ a set (sum of products) 
of smeared operators in region ${\cal R}'$, then we have 
\begin{equation}
0=\langle 0| O_{{\cal R}'}^1 O_{{\cal R}'}^2 \Phi(z)|0\rangle=\langle 0 | O_{{\cal R}'}^1  \Phi(z)O_{{\cal R}'}^2 |0\rangle~,
\end{equation}
where we have used that the operator $O^2_{{\cal R}'}$ is spacelike separated from the region where $\Phi$ is localized in order to move it to the right of $\Phi$.  
However, according to the Reeh-Schlieder theorem \cite{RS}, any state $|\psi\rangle$ may be approximated to arbitrary precision by acting with a (smeared) set of local operators in any open set in spacetime.   Since the region of points that are spacelike separated from a finite compact region 
is an open set,  we may apply this idea to the right-hand side above to conclude that for any sates $|\psi_{i}\rangle$

\begin{equation}
\langle\psi_{2}|\Phi|\psi_{1}\rangle=0~.
\end{equation}
This implies that $\Phi$ vanishes as an operator.

It is interesting to pinpoint exactly where this logic breaks down for non-local operators such as the average null energy operator $\mathcal{E}$. As long as the region that is spacelike separated to $\Phi$ is open, one may repeat the Reeh-Schlieder argument and prove that $\Phi$ vanishes even if it is non-local.  The way the null energy operator $\mathcal{E}$ avoids this conclusion is that it has support along a complete null line and hence the region of points that are spacelike separated to $\mathcal{E}$ is not open, since it consists just of the codimension one null plane containing the null line.\footnote{We that H. Casini for an enlightening discussion on this point.}

\section{Details of the Collider Calculation}\label{app:TTO}
In this appendix we will give more details on the calculation of the energy expectation value for a conformal collider experiment that we considered in this paper, giving a bound on $C_{TT\OO}$ in arbitrary dimensions.  

\subsection{Normalized States}
The states that we consider for the collider experiment are superposition of states of normalized wavepackets. Following \cite{Hofman:2008ar} we take the state defined as 
\beq
|\OO(q,\lambda)\rb \spc \equiv \spc \mathcal{N} \int d^dx~e^{-i q x^0} \spc \exp\left[{- \frac{x_0^2 + \vec{x}^2}{\sigma^2}} \right]\spc \lambda\cdot \OO (x) | 0\rb, ~~~ q>0
\eeq
where $q \sigma \gg 1$. We find the normalization $\mathcal{N}$ by requiring the state to have unit norm in this limit. We will give their values only for the operators and polarizations relevant to computing the bound on $C_{TT\OO}$. Namely for a scalar operator and for the stress tensor with polarization which is scalar with respect to the $SO(d-2)$ symmetry perpendicular to $n$.

We normalize the scalar operator such that its two-point function is $\lb \OO(x) \OO(0) \rb = C_\OO x^{-2\D}$. Then the normalization condition for the state considered in the collider experiment is
\beq\label{eq:norm0}
\lb \OO(q) | \OO(q) \rb =1 ~~\Rightarrow~~\mathcal{N}_\OO^{-2} =C_\OO \frac{2 \pi^{\frac{d+2}{2}}}{\Gamma(\Delta)\Gamma(\Delta-\frac{d}{2}+1)} \left( \frac{q}{2} \right)^{2\Delta-d}
\eeq
where we used the following integral identity
\beq
\int d^d x~ \frac{e^{i q x^0}}{x^{2\D}} = \frac{2 \pi^{\frac{d+2}{2}}}{\Gamma(\D)\Gamma(\D -\frac{d}{2}+1)} \left( \frac{q}{2} \right)^{2\D - d},~~~q>0
\eeq

We will also need the proper normalization for the scalar state created by the stress tensor, which has the form 
\beq
| T (q, \lambda_0) \rb \spc \equiv \spc  \mathcal{N}_{T_0}\spc \int d^d x \spc e^{- i q x^0} \spc (\lambda_0)_{ij} T_{ij}(x) | 0\rb,
\eeq
where we assumed the localized wavepacket limit. If we use conservation of the stress tensor we can chose the polarization along spatial directions. The normalized scalar polarization is
\beq\label{def:scalarpol}
(\lambda_0)_{ij} = \sqrt{\frac{d-1}{d-2}} \left[  \spc n_i n_j-\frac{1}{ (d-1)} \spc\delta_{ij} \right],
\eeq
which satisfies ${\rm Tr}(\lambda_0)=0$ and $\lambda_0 \cdot \lambda_0 = 1$. In this case the normalization condition gives 
\beq\label{eq:normT0}
\lb T(q,\lambda_0) | T(q,\lambda_0)\rb = 1 ~~\Rightarrow ~~ \mathcal{N}_{T_0}^{-1/2} =C_T \frac{4 (d-1) \pi ^{\frac{d}{2}+1}}{\Gamma \left(\frac{d}{2}\right) \Gamma (d+2)} \left(\frac{q}{2}\right)^d,
\eeq
where the normalization of the two-point function is 
\beq
\lb T_{\mu\nu}(x) T_{\rho\sigma}(0) \rb = \frac{C_T}{x^{2d}}\spc I_{\mu\nu\rho\sigma}(x), \label{CTdef}
\eeq
and the tensor structure that appears derived in \cite{Osborn:1993cr} is
\bea
I_{\mu\nu\rho\sigma}(x) &=& \frac{I_{\mu\rho}(x)I_{\nu\sigma}(x)+I_{\mu\sigma}(x)I_{\nu\rho}(x)}{2}- \frac{1}{d}\spc g_{\mu\nu}\spc g_{\rho\sigma},\\
I_{\mu\nu}(x) &=& g_{\mu\nu} - 2 \frac{x_\mu x_\nu}{x^2} \label{eq:vectorIA} .
\ea
Of course, by $SO(d-1)$ rotational symmetry, the normalizations for $T_1$ and $T_2$ are also given by 
\nref{eq:normT0}, once the polarizations are normalized to unity. Below we will perform the experiment of \cite{Hofman:2008ar} for linear superpositions of these normalized states. But first we will review the form of the correlators we will need, mainly to fix notation and conventions.

\subsection{Three-Point Functions}
The three-point functions we will need are $\lb TTT\rb$, $\lb T\OO \OO\rb$ and $\lb TT\OO\rb$. Their form were derived in \cite{Osborn:1993cr} and the first two were studied in the context of the conformal collider in four dimensions in \cite{Hofman:2008ar} and generalized to arbitrary dimensions in \cite{Buchel:2009sk}. First, we will focus on $\lb T T\OO \rb$ which was not studied previously in the context of the conformal collider. The form consistent with conformal symmetry and conservation of the stress-tensor found in \cite{Osborn:1993cr} is
\beq
\lb T_{\mu\nu} (x_1) T_{\rho\sigma}(x_2) \OO(x_3) \rb = \frac{1}{x_{12}^{2d-\Delta} x_{23}^\Delta x_{31}^\Delta} I_{\mu\nu}{}^{\mu' \nu'}(x_{13}) I_{\rho\sigma}{}^{\rho' \sigma'}(x_{23}) t_{\mu' \nu' \rho' \sigma'}(X_{12}), 
\eeq
where $X_{12} = \frac{x_{13}}{x_{13}^2}-\frac{x_{23}}{x_{23}^2}$ and the tensor structure is a sum of three terms  
\beq
t_{\mu \nu \rho \sigma}(x) \equiv  \hat{a} \spc h^1_{\mu \nu \rho \sigma}(x) + \hat{b} \spc h^2_{\mu \nu \rho \sigma} (x) + \hat{c}\spc h^3_{\mu \nu \rho \sigma}(x),
\eeq
where each $h^i$ is traceless and symmetric under $\mu\nu \leftrightarrow \rho\sigma$, $x\to-x$ 
\bea
h^1_{\mu \nu \rho \sigma}(x)& =&\frac{1}{x^4}(x_\mu x_\nu x_\rho x_\sigma + \ldots),\\
h^2_{\mu \nu \rho \sigma}(x) &= &\frac{1}{x^2}(x_\mu x_\rho g_{\nu\sigma} + \ldots), \\
h^3_{\mu \nu \rho \sigma}(x)& =& g_{\mu\rho} \spc g_{\nu\sigma} +\ldots, 
\ea
where dots represent terms needed for expressions to be traceless and symmetric and in each line they involve a fixed number of factors of $x$. The main advantage of this approach is that it makes transparent the OPE limit $x_2\to x_1$, or equivalently taking the $x_3 \to \infty$ limit 
\beq
T_{\mu\nu}(x) T_{\rho\sigma}(0) \sim \frac{1}{x^{2d-\Delta}}(\hat{a} \spc h^1_{\mu \nu \rho \sigma}(x) + \hat{b} \spc h^2_{\mu \nu \rho \sigma} (x) + \hat{c} \spc h^3_{\mu \nu \rho \sigma}(x)) \mathcal{O}(0).
\eeq
Conservation can be imposed in this limit to the right hand side to the equation above, giving the two independent relations 
\bea\label{eq:conservationTTO}
\hat{a}+4\hat{b} - \frac{1}{2} (d-\Delta) (d-1)(\hat{a}+4\hat{b})-d \Delta \hat{b} &=&0,\\
\hat{a}+4\hat{b}+d(d-\Delta)\hat{b} + d(2d-\Delta) \hat{c} &=&0.
\ea
This fixes the three point function to a single conserved structure up to an overall coefficient, which we can define as 
\beq \la{OPEcoD}
C_{TT\OO} \spc \equiv \spc  \hat{a}+8(\hat{b}+\hat{c}). 
\eeq

Another standard way of representing conformal three-point functions is given by the spinning correlator formalism of \cite{Costa:2011mg}. We will write the correlator in terms of the embedding space coordinate $X_i\in \mathbb{R}^{d+1,1}$ and the polarization $Z_i \in \mathbb{R}^{d+1,1}$ such that $Z^2=0$. The correlator we need is in appendix A of \cite{Costa:2011mg} and in terms of the usual conformal structures $V_i$ and $H_{ij}$ is given by
\beq\label{eq:TTO}
\lb T(X_1,Z_1) T(X_2,Z_2) \OO(X_3)\rb = \frac{\alpha_1 V_1^2 V_2^2 + \alpha_3 H_{12} V_1 V_2 + \alpha_6 H_{12}^2 }{(-2X_1\cdot X_2)^{d+2 - \frac{\Delta}{2}} (-2X_2\cdot X_3)^{\frac{\Delta}{2}}(-2X_3\cdot X_1)^{\frac{\Delta}{2}}},
\eeq
where $T(X,Z)\equiv Z^A Z^B T_{AB}(X)$ (with the index running from $0$ to $d+1$) and the labels $\alpha$ corresponds to the subset of the 10 structures that $\lb TT\mathcal{O}_\ell \rb$ has when $\ell=0$. The building blocks are
\bea
V_1 &=&\frac{ (Z_1 \cdot X_2) (X_1 \cdot X_3) -(Z_1 \cdot X_3) (X_1 \cdot X_2) }{X_2\cdot X_3}\\
V_2 &=&\frac{ (Z_1 \cdot X_3) (X_2 \cdot X_1) -(Z_1 \cdot X_1) (X_2 \cdot X_3) }{X_1\cdot X_3}\\
H_{12} &=& -2((Z_1\cdot Z_2)(X_1\cdot X_2) -(Z_1 \cdot X_2)(Z_2 \cdot X_1)).
\ea
Starting from the expression in $d+2$-dimensional embedding space we can obtain the $d$-dimensional correlator $\lb T(x_1,z_1) T(x_2,z_2) \OO(x_3) \rb$ by using the replacements $-2 X_i\cdot X_j\to x_{ij}^2$, $Z_i \cdot Z_j\to z_i \cdot z_j$ and $X_i \cdot Z_j \to x_{ij} \cdot z_j$, where now in $d$-dimensions we define $T(x,z)= z^\mu z^\nu T_{\mu\nu}(x)$, with the index running from $0$ to $d-1$. Of course after these replacements the answer coincides with the Osborn-Petkou three-point function. We can match the coefficients of the different representations by taking the OPE limit. The result gives 
\bea\label{eq:OPtoSC}
\alpha_1 &=& \hat{a} + 8(\hat{b}+\hat{c}),\\
\alpha_3 &=& 4(\hat{b}+2\hat{c}),\\
\alpha_6 &=& 2 \hat{c},
\ea
The OPE coefficient  \nref{OPEcoD} is now $C_{TT\OO} = \hat{a} + 8(\hat{b}+\hat{c}) = \alpha_1$. The conservation equations in terms of these parameters are 
\bea
&& \alpha_1(2 + \Delta - d(1-d +\Delta))+\alpha_3 \Big(-2 - \Delta + \frac{d}{2}(\Delta+2)\Big)=0\\
&&2 \alpha_1 + \frac{1}{2} \alpha_3 (-4 + d^2 - d \Delta) + \alpha_6 d \Delta =0.
\ea
which we can solve in terms of $C_{TT\OO}$. 

We defined in the main text the notation we will use for the stress-tensor three-point function. Another correlator we need is 
\beq
\lb T_{\mu\nu}(x_1) \OO(x_2) \OO(x_3) \rb = C_{T\OO \OO} \frac{1}{x_{12}^d x_{23}^{2\Delta-d} x_{31}^d}I_{\mu\nu\rho\sigma}(x_{13}) \left(\frac{X^\rho_{12} X^\sigma_{12}}{X_{12}^2} - \frac{1}{d} g^{\rho\sigma}\right),
\eeq
which is fixed by a Ward identity to be $C_{T\OO \OO} = - C_\OO \frac{d  \Delta}{(d-1) \Omega_{d-1}}$, with $\Omega_{d}$ being the are of a $S^d$ sphere. 

\subsection{Energy Matrix}
As explained in the main text we want to consider states of the form 
\beq
| \Psi \rb = v_1 | T(q,\lambda_0) \rb + v_2 | \OO(q)\rb,
\eeq
where we take $v=(v_1,v_2) \in \mathbb{C}^2$ such that $|v_1|^2 + |v_2|^2 =1$. The energy one-point function in the collider experiment is 
\beq
\lb \Psi| \mathcal{E}(n) | \Psi \rb = v^\dagger \left(\begin{array}{cc} \lb T(q,\lambda_0) | \mathcal{E}(n) | T(q,\lambda_0)\rb & \lb T(q,\lambda_0) | \mathcal{E}(n) | \OO(q)\rb  \\ \lb T(q,\lambda_0) | \mathcal{E}(n) | \OO(q)\rb^* &\lb \OO(q) | \mathcal{E}(n) | \OO(q) \rb  \end{array}\right) v 
\eeq
In this section we will compute the entries of this matrix. The diagonal elements were already computed in \cite{Hofman:2008ar, Buchel:2009sk} and are given by 
\beq
\lb T(q,\lambda_0) | \mathcal{E}(n) | T(q,\lambda_0)\rb  = \frac{q}{\Omega_{d-2}} \left( 1 -\frac{d-3}{d-1} t_2 -\frac{d(d-1)-4}{d^2-1} t_4\right),
\eeq
where
\begin{equation}
t_2 =\frac{\left(d^2-1\right) ((d-3) N_F-2 (d-2) N_V)}{(d-3) ((d-1) (d N_V+N_F)+2N_B)}~, \hspace{.3in}t_4=\frac{\left(d^2-1\right) (N_B-N_F+N_V)}{2 N_B+(d-1) (d N_V+N_F)}~.
\end{equation}
Using these expressions we can find the parameters we called $T_{0}$, $T_1$ and $T_2$ in the main text. They are given by equation \eqref{Tsdef} where the functions $\rho_i(d)$ are given by 
\bea\label{appalphas}
\rho_0(d) &=&\frac{1}{\Omega_{d-1}^2}\frac{d(d+1)(d-2)}{2(d-1)} \\
\rho_1(d) &=& \frac{1}{\Omega_{d-1}^2}\frac{d(d+1)}{4} \\
\rho_2(d) &=& \frac{1}{\Omega_{d-1}^2}\frac{d(d+1)(d-2)}{2(d-3)}
\ea

The state created by a scalar operator gives
\beq
\lb \OO(q) | \mathcal{E}(n) | \OO(q)\rb=\frac{q}{\Omega_{d-2}}.
\eeq

Now we will obtain the off-diagonal element of this matrix we has not been computed in the literature. To perform the calculation in arbitrary dimensions it is convenient to use the spinning correlator formalism. Since we are computing an expectation value the correlator we need to consider is not time-ordered. The right $i \epsilon$ prescription for this purpose was explained in \cite{Hofman:2008ar} and \cite{Buchel:2009sk}, and we will omit it here to ease the notation. We start from the $d$-dimensional expression
\beq
\lb T(x_1,z_1)T(x_2,z_2) \OO(x_3) \rb = \frac{\alpha_1 V_1^2 V_2^2 + \alpha_3 H_{12} V_1 V_2 + \alpha_6 H_{12}^2 }{(x_{12}^2)^{d+2 - \frac{\Delta}{2}} (x_{23}^2)^{\frac{\Delta}{2}}(x_{31}^2)^{\frac{\Delta}{2}}}
\eeq
We will chose $T(x_2,z_2)$ to be the insertion taken to infinity and giving $\mathcal{E}(n)$. First we take $z_2 =m=(1,n)$ (we chose the mostly plus convention for the metric in Minkowski space). Therefore $T(x_2,z_2) \to \frac{1}{4} T_{--}(x_2)$. Then we take the limit 
\beq
x_2 \cdot \bar{m} \to \infty,~~~~\bar{m} = (-1,n).
\eeq
To take this limit we can use the results of appendix F of  \cite{Komargodski:2016gci}, and obtain 
\beq
\lim_{x_2\cdot \bar{n}\to\infty} \lb T(x_1,z_1) r^{d-2}T_{--}(x_2) \OO(x_3)\rb =  \frac{\alpha_1 \hat{V}_1^2 \hat{V}_2^2 + \alpha_3 \hat{H}_{12} \hat{V}_1 \hat{V}_2 + \alpha_6 \hat{H}_{12}^2 }{2^d (x_{12}\cdot m)^{d+2 - \frac{\Delta}{2}} (x_1^2)^{\frac{\Delta}{2}}(x_2\cdot m)^{\frac{\Delta}{2}}}
\eeq
where the structures in this limit are 
\beq
\hat{V}_1 = \frac{z_1 \cdot m \frac{x_1^2}{2}-x_1\cdot z_1 x_{12}\cdot m}{x_2\cdot m},~~~\hat{V}_2 = \frac{x_1 \cdot m}{x_1^2},~~H_{12}= - z_1 \cdot m.
\eeq
We can get the correct polarization of the insertion $T(x_1,z_1)$ by replacing $z_\mu z_\nu \to \lambda^T_{\mu\nu}$, assuming $\lambda$ is already traceless and symmetric which is true for expression \eqref{def:scalarpol}. To simplify the expressions we will choose $n=(1,0,\ldots,0)$ and write the positions as $x=(x^+,x^-, x^\perp)$, where $x^+=x\cdot \bar{m} = x^1 + x^0$, $x^-=x\cdot m = x^1 - x^0$ and $x^\perp$ corresponds to the $d-2$ transversal components. Then we can define
\beq
\hat{G} =\lim_{x_2^+\to\infty}(x_2^+/2)^{d-2} \lb \lambda_0 \cdot T(x_1) T_{--} (x_2^+,x_2^-,0) \OO(x_3=0)\rb 
\eeq
which is given by 
\bea
\hat{G} &=& \alpha_1\lambda^T_{11}\frac{(x_1^-)^2( \frac{x_1^4}{4}-x_1^1 x_{12}^- x_1^2+ x_1^1 x_1^1 (x_{12}^-)^2-\frac{1}{d-2}\sum_{i_\perp}(x_1^{i_\perp})^2 (x_{12}^-)^2)}{2^{d}(x_{12}^-)^{d+2 - \frac{\Delta}{2}} (x_1^2)^{\frac{\Delta}{2}+2}(x_2^-)^{\frac{\Delta}{2}+2}}\nn
&&+ \alpha_3 \lambda^T_{11}\frac{2^{-d}(x_1^-)(x_1^1 x_{12}^- - \frac{1}{2} x_1^2)}{(x_{12}^-)^{d+2 - \frac{\Delta}{2}} (x_1^2)^{\frac{\Delta}{2}+1}(x_2^-)^{\frac{\Delta}{2}+1}}+\alpha_6 \lambda^T_{11} \frac{2^{-d}}{(x_{12}^-)^{d+2 - \frac{\Delta}{2}} (x_1^2)^{\frac{\Delta}{2}}(x_2^-)^{\frac{\Delta}{2}}}
\ea

First we do integral over $x_2^-$, using the following identity
\beq
\int dx_2^- \frac{1}{(x_2^- - i \epsilon)^{b}(x_{12}^- - i \epsilon)^{a}} = \frac{2 \pi i}{(x_1^- -2 i \epsilon)^{a+b-1}}\frac{\Gamma(a+b-1)}{\Gamma(a)\Gamma(b)} ~,
\eeq
 where we made explicit the pole prescription.
Finally, to take the limit of the localized wavepackets is equivalent to setting $x_3 \to 0$ and make a Fourier transform with respect to $x_1$ with momentum $(q,0,\ldots,0)$, namely
\beq
\int d^d x_1 e^{-i q x_1^0} \int dx_2^- \hat{G}.
\eeq
To do this we first integrate over the $d-2$ transverse directions $x_1^\perp$ and then integrate over the light-cone coordinates $x_1^\pm$. Because of $SO(d-2)$ invariance, the integrand only depends on $x_1^+$, $x_1^-$ and $x_1^\perp \cdot x_1^\perp$. Then the integral can be written as
\bea
\int d^d x_1 e^{-i q x_1^0} \int dx_2^- \hat{G}&=&\frac{\Omega_{d-3}}{2} \int dx_1^+ e^{- i \frac{q}{2} x_1^+} \int dx_1^- e^{i \frac{q}{2} x_1^-} ,\nn
&&\times\int R^{d-3}dR  \spc F(x_1^+,x_1^-,(x_1^\perp)^2 = R^2),
\ea
where we defined $F=\int dx_2^- \hat{G}$ to indicate the functional dependence explicitly. After performing these integrals we use conservation conditions to write $\alpha_3$ and $\alpha_6$ in terms of $\alpha_1 = C_{TT\OO}$ using equations \eqref{eq:conservationTTO} and \eqref{eq:OPtoSC}. Combining the three structures gives 
\beq
\int d^d x_1 e^{-i q x_1^0} \int dx_2^- \hat{G} =  \frac{ C_{TT\OO} 2^{2-d}(d-1)  \pi ^{\frac{d}{2}+2} \Gamma (d+1)}{(d-2) \Gamma \left(\frac{\Delta }{2}+2\right)^2 \Gamma \left(d-\frac{\Delta }{2}\right) \Gamma \left(\frac{d+\Delta }{2}\right)} \lambda_0 .n.n\spc \left(\frac{q}{2} \right)^{\Delta+1}
\eeq
In this expression we generalized the answer to arbitrary $n$ by replacing $\lambda_{11} \to \lambda_{ij} n^i n^j$. Finally, we need to replace the specific value of the polarization tensor \eqref{def:scalarpol} and the proper normalization of the collider states \eqref{eq:normT0} and \eqref{eq:norm0}. The final answer for the off-diagonal entry of the energy matrix is 
\beq
\lb T(q,\lambda_0) | \mathcal{E}(n) | \OO(q)\rb = \frac{q}{\Omega_{d-2}}\frac{C_{TT\OO}}{\sqrt{C_T C_\OO}} h(\Delta), 
\eeq 
where 
\beq\label{eq:defH}
h(\D ) \equiv  \frac{ \pi ^{\frac{d+1}{2}} \Gamma (d+1) \sqrt{\Gamma \left(\frac{d}{2}-1\right)\Gamma (d+2)\Gamma (\Delta )\Gamma \left(\Delta-\frac{d-2}{2}\right)}}{2^{d}\Gamma \left(\frac{d-1}{2}\right) \Gamma \left(\frac{\Delta }{2}+2\right)^2  \Gamma \left(\frac{d+\Delta }{2}\right)\Gamma \left(d-\frac{\Delta }{2}\right)}
\eeq
Then the energy matrix that gives the expectation value for these superposition states is 
\beajm
\lb \Psi| \mathcal{E}(n) | \Psi \rb & =&  v^\dagger \left(\begin{array}{cc} T_0&\frac{C_{TT\OO}}{\sqrt{C_T C_\OO}} h(\Delta)  \\ \frac{C_{TT\OO}^*}{\sqrt{C_T C_\OO}} h^*(\Delta) & 1 \end{array}\right) v 
\\
T_0 & \equiv & 1 -\frac{d-3}{d-1} t_2 -\frac{d(d-1)-4}{d^2-1} t_4  
\eeajm

Having computed the energy matrix the next step is to impose ANEC, which is equivalent to imposing positivity of the energy expectation value for the collider experiment. This means that for all states  
\beq\label{eq:stateA}
|\Psi(v) \rb = v_1 |  T(q,\lambda_0) \rb + v_2 | \OO(q)\rb
\eeq
we need to impose 
\beq
\lb \Psi (v) | \mathcal{E}(n) | \Psi (v)\rb >0 ,~~~\forall v\in \mathbb{C}^2.
\eeq
This constraint is equivalent to the positivity of all the leading principal minors of the energy matrix. The first constraint is 
\beq\label{eq:TTTboundA}
T_0 = 1 -\frac{d-3}{d-1} t_2 -\frac{d(d-1)-4}{d^2-1} t_4 = \rho_0(d) \left(\frac{N_B}{C_T}\right)\geq 0
\eeq
which is the same as the original constraints of \cite{Hofman:2008ar} and \cite{Buchel:2009sk}. The next minor imposes the positivity of the 2$\times$ 2 matrix which is 
\beq\label{eq:TTOboundA}
\frac{|C_{TT\OO}|^2}{C_TC_\OO} |h(\Delta)|^2 \leq T_0
\eeq
We can write $T_0$ and $C_T$ in terms of the $\lb TTT\rb$ structures $N_B$, $N_F$ and $N_V$. This gives the equivalent expression that we quoted in the introduction 
\beq
\frac{|C_{TT\OO}|^2}{C_\OO} f(\Delta) \leq N_B
\eeq
where 
\beq
f(\Delta) = \frac{(d-1)^3 d \pi ^{2 d} \Gamma \left(\frac{d}{2}\right) \Gamma (d+1) \Gamma (\Delta ) \Gamma \left(\Delta -\frac{d-2}{2}\right)}{(d-2)^2 \Gamma \left(\frac{\Delta }{2}+2\right)^4  \Gamma \left(\frac{d+\Delta }{2}\right)^2\Gamma \left(d-\frac{\Delta }{2}\right)^2}
\eeq
These two conditions \eqref{eq:TTTboundA} and \eqref{eq:TTOboundA} are necessary and sufficient for the energy to be positive for any state of the form \eqref{eq:stateA}. For operators $\OO$ that are not hermitian this bound does not have information about the phase of the OPE coefficient $C_{TT\OO}$. 

\section{Free Scalar Correlators}
In this section we will present some details on the calculation of the $TT\OO$ correlators for a free scalar that saturates the bound above. We use the normalization of \cite{Osborn:1993cr} for
\beq
\lb \phi(x) \phi(0) \rb = \frac{1}{(d-2)\Omega_{d-1}} \frac{1}{x^{d-2}}.
\eeq
and the stress tensor is defined as 
\beq
T_{\mu\nu} = \spc :\hspace{-0.8mm}\partial_\mu \phi \partial_\nu \phi \hspace{-1mm}: - \frac{1}{4(d-1)}((d-2)\partial_\mu\partial_\nu + g_{\mu\nu}\partial^2):\hspace{-0.8mm}\phi^2\hspace{-1mm}:.
\eeq
Since scalar operators have integer dimensions we only need to consider $\OO$ such that $\Delta < 2d$. The first one is $\OO \sim \phi$. This one is predicted to vanish since $f(\Delta=\frac{d-2}{2}) \to \infty$. This is indeed the case since an odd number of fields appear in $\lb TT\phi\rb$. The next operator is $\OO = \spc :\hspace{-0.8mm}\phi^2\hspace{-1mm}:$ of dimension $\Delta= d-2$. The correct normalization of the two-point function gives 
\beq
\lb :\hspace{-0.8mm}\phi^2\hspace{-1mm}:(x):\hspace{-0.8mm}\phi^2\hspace{-1mm}:(0)\rb = \frac{2}{(d-2)^2 \Omega_{d-1}^2}\spc \frac{1}{x^{2(d-2)}},~~~C_{\OO}=\frac{2}{(d-2)^2 \Omega_{d-1}^2} .
\eeq
Using Wick contractions we can also compute $\lb TT\phi^2\rb$. One can check that the answer has the conformal invariant structure \eqref{eq:TTO} with
\beq
C_{TT\OO} = \alpha_1 = \frac{(d-2) d^2}{2 (d-1)^2}  \frac{1}{\Omega_{d-1}^3},~~~\alpha_3 =- \frac{4}{d-2}\alpha_1 ,~~~\alpha_6=\frac{2}{(d-2)d}\alpha_1
\eeq
Finally, the function appearing in the bound takes the value 
\beq
f(\Delta = d-2) =\frac{8 (d-1)^4 \pi ^{2 d}}{(d-2)^4 \Gamma \left(\frac{d}{2}+1\right)^4}
\eeq
Putting everything together we find that the bound is saturated
\beq
\frac{|C_{TT\OO}|^2}{C_{\OO}} f(d-2)=1 \leq N_B = 1.
\eeq
 
We have seen that the bound is saturated by a scalar field with $\OO=\spc :\hspace{-0.8mm}\phi^2\hspace{-1mm}:$. Nevertheless there is one more primary scalar field we can make of dimension less than $2d$, namely $\OO = \spc :\hspace{-0.8mm}\phi^4\hspace{-1mm}:$ which has dimension $\Delta_{\phi^4}=2(d-2)$. Working out the Wick contractions we can verify that $\lb TT :\hspace{-0.8mm}\phi^4\hspace{-1mm}: \rb =0$.
We can argue more generally that this is so. The form of the correlator \eqref{eq:TTO} indicates that there is a $x_1 \to x_2$ singularity whenever $\Delta < 2d$. On the other hand, to have a non-zero answer we can only take a Wick contraction which is between $T$ and $:\hspace{-0.8mm}\phi^4\hspace{-1mm}:$ but not between the stress tensors. Therefore if this calculation would give a non-zero answer, it will be finite when $x_1 \to x_2$. The only way this is consistent with the form of the correlator fixed by conformal symmetry \eqref{eq:TTO} is if it indeed vanishes.

\section{$\lb TTO\rb$ Parity-Odd Structures in $d=3$}\label{app:d3odd}
In dimensions $d\geq4$ the three-point function $\lb TTO\rb$ has only a parity-even structure consistent with permutation symmetry and conservation of the stress-tensors. The situation for $d=3$ is special since only for this number of dimensions a new parity-odd structure appears, that is also consistent with all the requirements. In this case the full correlator is
\beq
\lb TT \OO \rb = \lb TT \OO \rb_{\rm even} +\lb TT \OO \rb_{\rm odd}
\eeq
where the parity-even part coincides with the answer for $d\geq4$ 
\beq
\lb T(X_1,Z_1) T(X_2,Z_2) \OO(X_3)\rb_{\rm even} = \frac{\alpha_1 V_1^2 V_2^2 + \alpha_3 H_{12} V_1 V_2 + \alpha_6 H_{12}^2 }{X_{12}^{5 - \frac{\Delta}{2}} X_{23}^{\frac{\Delta}{2}}X_{31}^{\frac{\Delta}{2}}},
\eeq
and the new structure is 
\beq
\lb T(X_1,Z_1) T(X_2,Z_2) \OO(X_3)\rb_{\rm odd} = \frac{\beta_1 V_1 V_2 + \beta_2 H_{12} }{X_{12}^{5 - \frac{\Delta}{2}} X_{23}^{\frac{\Delta}{2}}X_{31}^{\frac{\Delta}{2}}}\spc \epsilon(Z_1,Z_2,X_1,X_2,X_3),
\eeq
Since conservation put constraints independently for $\alpha_{1,3,6}$ and $\beta_{1,2}$ we can forget about the parity-even part and we get $\beta_1(\Delta-3) -\beta_2 (\Delta+1)=0$. Therefore the parity-odd structure is also fixed by a single OPE coefficient which we denote $C_{TTO}^{\rm odd}$, as opposed to the one in the even part $C_{TTO}^{\rm even} = \alpha_1$. For completeness we present the same correlator in the Osborn and Petkou formalism
\beq
\lb T_{\mu\nu} (x_1) T_{\rho\sigma}(x_2) \mathcal{O}(x_3) \rb = \frac{1}{x_{12}^{6-\Delta} x_{23}^\Delta x_{31}^\Delta} I_{\mu\nu}{}^{\mu' \nu'}(x_{13}) I_{\rho\sigma}{}^{ \rho' \sigma'}(x_{23}) t_{\mu' \nu' \rho' \sigma'}(X_{12}), 
\eeq
where $t= \hat{d}\spc  t^1 + \hat{e}\spc  t^2$ and we define 
\bea
t^1_{\mu\nu\rho\sigma}(X) &=& \frac{x_\mu x_\rho x_\gamma}{x^3} \varepsilon_{\nu \sigma \gamma} + \ldots,\\
t^2_{\mu\nu\rho\sigma}(X) &=& \frac{\delta_{\mu\rho} x_\gamma}{x} \varepsilon_{\nu \sigma\gamma} + \ldots,
\ea
where the dots represent other terms of the same form to make the answer symmetric, traceless and permutation symmetric. Conservation imposes $\hat{e}(\Delta-7)+\hat{d}(\Delta-3)=0$. Either in the Osborn and Petkou formalism or in the spinning correlator formalism, we define the parity-odd OPE coefficient as $C_{TTO}^{\rm odd} \equiv \hat{d}+\hat{e} = (\beta_2 - \beta_1)/4$. 

\section{$\lb TTJ\rb$ Three-Point Function}
\label{TTJapp}
In this appendix we will provide some details on the CFT three-point function controlling the mixed gauge-gravitational anomaly $\lb TTJ\rb$. Imposing permutation symmetry between the stress-tensors and conservation, this correlator only involves an allowed parity-odd structure. In the spinning correlator formalism it is given by 
\beq
\lb T(X_1,Z_1) T(X_2,Z_2) J(X_3,Z_3)\rb \sim \frac{H_{12}-4 V_1 V_2}{X_{12}^4 X_{23}^2 X_{31}^2} \spc \epsilon(Z_1,Z_2,Z_3, X_1,X_2,X_3),
\eeq 
where as usual the upper-case denote coordinates in embedding space and $H_{ij}$ and $V_i$ are the usual structures defined in \cite{Costa:2011mg}. From this expression it is possible to deduce the conservation equation for the current when the CFT is placed on a curved background and gives the right normalization for $C_{TTJ}$ \cite{Erdmenger:1999xx}. Then the three-point function is 
\beq
\lb T(X_1,Z_1) T(X_2,Z_2) J(X_3,Z_3)\rb = \frac{C_{TTJ}}{2\pi^6} \spc \frac{H_{12}-4 V_1 V_2}{X_{12}^4 X_{23}^2 X_{31}^2} \spc \epsilon(Z_1,Z_2,Z_3, X_1,X_2,X_3)
\eeq

For completeness we can write this same correlator using the notation of Osborn and Petkou. This can be written as 
\beq
\lb T_{\mu\nu} (x_1) T_{\rho\sigma}(x_2) J_\alpha(x_3) \rb = \frac{1}{x_{12}^{5} x_{23}^3 x_{31}^3} I_{\mu\nu}{}^{\mu' \nu'}(x_{13}) I_{\rho\sigma}{}^{\rho' \sigma'}(x_{23}) t_{\mu' \nu' \rho' \sigma' \alpha}(X_{12}) 
\eeq
where $t_{\mu \nu \rho \sigma \alpha}(x)$ is the OPE structure $T_{\mu \nu}(x)T_{\rho \sigma }(0)\sim |x|^{-5} t_{\mu \nu \rho \sigma \alpha}J^\alpha(0)$ and $X_{12} = \frac{x_{13}}{x_{13}^2} - \frac{x_{23}}{x_{23}^2}$. The two structures possible, which are linear combinations of the $H_{12}$ and $V_1V_2$, are explicitly 
\beq
t^1_{\mu \nu \rho \sigma \alpha}(x) = \frac{x_\gamma x_\mu x_\rho}{4|x|^3} \epsilon_{\nu \sigma \alpha \gamma} + \ldots,
\eeq
and 
\beq
t^2_{\mu \nu \rho \sigma \alpha}(x) = \frac{x_\gamma\delta_{\mu \rho}}{4|x|} \epsilon_{\nu \sigma \alpha\gamma}+\ldots,
\eeq
where the dots represent terms needed to add in ordered for the expression to be symmetric, traceless and permutation symmetric between the first two pair of indices. The most general case has $t=\hat{a} t^1 + \hat{b} t^2$. Imposing conservation and comparing with the spinning correlator formalism we get $\hat{a} =-6\hat{b}= 3C_{TTJ}/\pi^6$. Using this information it is straightforward to apply the same procedure as was done for $\lb TTO\rb$ to obtain the energy matrix elements in the conformal collider experiment.

   \section{Computing the Bound in the Gravity Theory } 
 \la{GravBound}
 
 In this appendix we relate the OPE coefficient $C_{TT{\cal O} } $ to a coefficient, $\alpha$,  in the $AdS_D$ effective action     
 \be \la{FullAct} 
 S = { M_{pl}^{D-2} \over 2 } \left[ \int \sqrt{g} ( R - 2 \Lambda ) + (\nabla \chi )^2 - m^2 \chi^2  + 2 \alpha \chi W^2 \right] ~,~~~~\Lambda = - { (D-1)(D-2) \over 2 R_{AdS}^2 }~,
 \ee
 where $D$ is the dimension of  $AdS_D$.  $\chi$ is defined to be dimensionless and $\alpha $ has dimensions of length squared. 

In principle we can compute the relation between $\alpha$ and $C_{TT{\cal O} } $ by computing the three point function between a scalar and the graviton produced
by this cubic term in the Lagrangian, using Witten diagrams. Instead,  we will follow a different route. We will directly compute the energy correlator in gravity and derive a bound on 
$\alpha$ by demanding its positivity. We then relate $\alpha$ and $C_{TT{\cal O }}$ by demanding that this gravity bound, in terms of $\alpha$, matches the 
bound we obtained in terms of $C_{TT {\cal O}}$ in the field theory analysis. 

 We will  rely on \cite{Hofman:2008ar,Buchel:2009sk} where the energy correlators were computed in gravity. 
 An important point is that the insertion of $T_{--}$ corresponds to a shock wave localized in a null plane. 
 Furthermore, an operator insertion at the origin with definite energy-momentum gives rise to an excitation that crosses this null plane 
 at a localized point. 
 For this reason the computation of the bound  boils down to analyzing the propagation of an excitation through a suitable gravitational shock wave in 
 flat space. The $AdS_D$ space is only relevant for determining the transverse profile of the shock wave, as we will see below. 
 
 For these reasons we consider a shock wave of the form 
  \be \la{metrPW}
  ds^2 = ds^2_{\rm flat} + (d x^+)^2 \delta(x^+) h(y)  ~,~~~~~~~ ds^2_{\rm flat} = - dx^+ dx^- + dy^2 ~.
  \ee
  Adding  gravitons we get 
  \be \la{PaGra}
  ds^2 = ds^2_{\rm flat} + (d x^+)^2 \delta(x^+) h(y) +  dx^\mu dx^\nu \zeta_{\mu } \zeta_\nu e^{ i p.x  }  G(p) +  
  dx^\mu dx^\nu \bar \zeta_{\mu } \bar \zeta_\nu e^{- i p.x }  \bar G(p)~,
  \ee 
  with  $\zeta^2=0$, $\zeta^\mu p_\mu =0$. Note that the graviton polarization is $\zeta_{\mu \nu} = \zeta_\mu \zeta_\nu $, 
  and is normalized to one $\zeta . \bar \zeta =1$.  
We can think of $G(p)$ and $\bar G(p)$ as complex numbers, which in the quantum theory will be related to $a$ and $a^\dagger$. 
 Inserting \nref{PaGra}   into \nref{FullAct} we can derive the   quadratic and cubic interaction terms. 
       \beajm \la{ExpaAc}
  S &=& {M_{pl}^{D-2} \over 2 }  \int dx^+ dx^- d^{D-2} y \left\{   \left[  p_+ p_- +  \delta(x^+) p_-^2 h \right] \left[ G(p) \bar G(p)  +  4 p_- p_+ \chi(p) \bar \chi(p) \right]+ 
  \right. 
  \cr
  & ~&  \left.
 +  8 p_-^2     \alpha     \zeta^{ij} \partial_i \partial_j h    \delta(x^+)   G(p) \bar \chi(p)   + c.c. \right\} ~,
  \eeajm
  where we only wrote the terms relevant for our computation, ignoring transverse derivatives in the kinetic terms. 
Momentarily setting the scalar field to zero, we see that we have the following equation for the graviton as it crosses the shock wave 
  \be
   \Delta h_{\mu \nu } \equiv h_{\mu \nu} |_{x^+ = 0^+} -h_{\mu \nu} |_{x^+ = 0^-} = i p_- h h_{\mu \nu } ~.
   \ee
  Exponentiating this, $h_{\mu \nu}(x^+ = 0^+) = e^{ i p_- h } h_{\mu \nu}(x^+= 0^-) $, we see that the 
   time delay is simply given by $h$. This is as expected from \nref{metrPW} since we can shift $x^-$ by $h$ and make the term involving $h$ disappear if
   we ignore its $y$ dependence.  
    So far, we considered the computation in flat space. An insertion 
    of the null energy integrated along a ray in the boundary theory 
    gives rise to a shockwave in $AdS_D$ which is localized on a null direction.  Its dependence on the transverse directions is the following. 
    The transverse space is an $H_{D-2}$. This is easy to see in embedding coordinates where $AdS_D$ is 
    $ \tilde W^+ \tilde W^- + W^\mu W_\nu = -1$ (setting $R_{AdS_D} =1$). The null plane is $\tilde W^+=0$.
     It contains the null direction parametrized by $\tilde W^-$ as 
    well as the transverse space $W^\mu W_\mu =-1$. 
    The profile of the wave is proportional to $ h \propto  ( W^0 - W^i n^i )^{2-D}$  \cite{Hofman:2008ar,Buchel:2009sk}, with a positive coefficient. 
    Here $\vec n^i$ is a vector on the sphere at infinity in the boundary Minkowski space. 
    For \nref{ExpaAc} we need the derivatives at $W^i=0$, which are given by  
   \be
   h \to h ~,~~~~~~ \partial_i \partial_j h =[ ({\rm constant})    \delta_{ij} +  (D-2) (D-1) n_i n_j ]  h { 1\over R^2_{AdS_D}}  ~,
   \ee
   where the constant does not matter because the graviton is traceless. 
   The relevant component of the graviton is the one with polarization along $n^i$. This has the expression
   \be \la{LoPo}
   \zeta_{ij} = \sqrt{ D-2 \over D-3} \left[ n^i n^j - { \delta_{ij} \over D-2} \right] ~.
   \ee
 The expression for the time delay acting on  a superposition of a graviton and a scalar is now a  matrix proportional to  
  \be \la{Gadef} \left( 
    \begin{array}{cc}  1 & \gamma 
    \\
  \gamma   & 1
     \end{array} \right)  ~,~~~\gamma  \equiv 4  (D-1) \sqrt{(D-3)(D-2) } { \alpha \over R_{AdS_D}^2} ~,
     \ee
     where the matrix is acting on a two dimensional space where one direction is the scalar and the other is the graviton with polarization \nref{LoPo}. 
The unitarity bound comes from the restriction that the eigenvalues are non-negative, or $|\gamma| \leq 1$, 
which is 
\be
 { |\alpha| \over R_{AdS_D}^2 }   \leq { 1 \over 4     (D-1) \sqrt{(D-3)(D-2) } } =  { 1 \over 4    d \sqrt{(d-2)(d-1) } } ~,
 \ee
   where $d$ is the dimension of the boundary. Comparing this with the bound obtained in \nref{BoundFT}, with the non-Einstein-gravity structures set to zero, we 
   obtain \nref{boundco}. Of course, once we get the proportionality constant between $\alpha$ and $C_{TT {\cal O}}$ for the Einstein gravity case, the 
   same constant holds also if we add the  purely gravitational higher derivative terms that generate the other tensor structures for $\langle TTT\rangle$. 
We could add them to this computation, but   we expect to reproduce the bounds we got in the general field theory analysis.

   \subsection{Four-Dimensional Case }
   
         In the special case of the four dimensional theory, we actually have two couplings  \nref{GravFd}. This leads to a new interaction term in 
   \nref{ExpaAc} of the form 
   \be
   \alpha     \zeta^{ij} \partial_i \partial_j h  \rightarrow   \alpha_e     \zeta^{ij} \partial_i \partial_j h   + \alpha_o  \zeta^{il } \epsilon_{lj }  \partial_i \partial_j h~,
   \ee
   where now $\epsilon_{ij}$ is the two dimensional epsilon symbol. This means that the scalar can now mix with the other graviton polarization 
   component besides \nref{LoPo}. Namely, defining \nref{LoPo} as  $\zeta_{\oplus} $, it can also mix with    $\zeta_{\otimes}^{ij} \equiv \epsilon^{i l} \zeta^{lj}_{\oplus} $. 
   Now the time delay is a a three by three matrix 
   \be \la{GaBedef} \left( 
    \begin{array}{ccc}  1 & \gamma &\beta
    \\
  \gamma   & 1 & 0 
  \\
  \beta & 0 & 1 
     \end{array} \right)  ~,~~~~~\gamma  \equiv 12  \sqrt{ 2  } { \alpha_e \over R_{AdS_4}^2}  ~,~~~~~\beta  \equiv 12  \sqrt{ 2  } { \alpha_o \over R_{AdS_4}^2}~,
     \ee
    where the rows and columns correspond to the scalar and the two graviton polarizations. 
     Now  the bound is \nref{AdSBound}. Comparing this to \eqref{eq:3d3x3b},  after setting the non-Einstein-gravity structures to zero, we get the precise mapping to 
     the $C_{TT{\cal O}}$ coefficients \nref{fdCoef}.

\end{appendix}

\bibliographystyle{utphys}
\bibliography{OPEbound}{}

\end{document}